\documentclass[twocolumn, aps, prb]{revtex4-2}

\usepackage{graphicx}
\usepackage{amsfonts}

\usepackage{amssymb}
\usepackage{amsmath}
\usepackage{type1cm}

\usepackage{multirow} 
\usepackage{tabularx} 
\usepackage{array}

\usepackage{braket}

\usepackage{bm}
\usepackage{color}

\usepackage{amsthm}

\allowdisplaybreaks[4]
\usepackage[margin=2cm]{geometry}
\usepackage{enumerate}

\begin{document}

\title{\bf Synchronization and chaos in a spin-torque oscillator with a perpendicularly magnetized free layer}
\author{Terufumi Yamaguchi$^{1}$} 
\email{yamaguchi-terufumi@aist.go.jp}
\author{Nozomi Akashi$^{2}$} 
\author{Kohei Nakajima$^{2}$} 
\author{Sumito Tsunegi$^{1}$}
\author{Hitoshi Kubota$^{1}$} 
\author{Tomohiro Taniguchi$^{1}$}
\email{tomohiro-taniguchi@aist.go.jp}

\affiliation{$^{1}$National Institute of Advanced Industrial Science and Technology (AIST), 
Spintronics Research Center, Tsukuba, Ibaraki 305-8568, Japan \\
$^{2}$Graduate School of Information Science and Technology, The University of Tokyo, Bunkyo-ku, Tokyo 113-8656, Japan}
\date{\today}

\begin{abstract}
  Synchronization and chaos caused by alternating current and microwave field in a spin torque oscillator consisting of 
  a perpendicularly magnetized free layer and an in-plane magnetized reference layer is comprehensively studied theoretically. 
  A forced synchronization by the alternating current is observed in numerical simulation over wide ranges of its amplitude and frequency. 
  An analytical theory clarifies that the nonlinear frequency shift, as well as the spin-transfer torque asymmetry, plays a key role 
  in determining locking range and phase difference between the oscillator and current. 
  Chaos caused by the alternating current is identified for a region of large alternating current by evaluating the Lyapunov exponent. 
  Similar results are also obtained for microwave field, 
  although the parameter regions causing chaos are narrower than those by the alternating current. 
\end{abstract}

\maketitle


\section{Introduction}
\label{sec:Introduction}

An excitation of a limit cycle oscillation of magnetization in a spin torque oscillator (STO) 
via spin-transfer torque effect \cite{slonczewski96,berger96} has attracted much attention \cite{kiselev03,rippard04,krivorotov05,bertotti05,houssameddine07,bertotti09text,rippard10,grimaldi14,tsunegi16}. 
This is because it provides an interesting example of nonlinear oscillator in nanoscale 
and therefore bridges between condensed matter physics and nonlinear science. 
In particular, synchronization to an external oscillating signal, such as an alternating current or microwave field, 
is of great interest because of it has wide locking range originating from the nonlinearity of the magnetization dynamics \cite{slavin09}. 
The injection of the time-dependent signal to an STO also leads to a fascinating phenomenon, namely chaos. 
The synchronization \cite{rippard05,zhou08,georges08,urazhdin10,finocchio12} and chaos \cite{li06,yang07,xu08} in STO have been extensively studied 
both experimentally and theoretically in previous works. 
We notice here that previous works mostly focused on an STO consisting of an in-plane magnetized free layer 
because STO as such was easy to fabricate. 


Recent development of the fabrication technology, however, enables us to develop STO with perpendicularly magnetized free layer \cite{zeng12,kubota13}. 
This achievement mainly is due to the discovery of the interfacial perpendicular magnetic anisotropy between CoFeB and MgO in a magnetic tunnel junction \cite{yakata09,ikeda10,kubota12}, 
which nowadays are commonly used in current spintronics devices. 
A large emission power with narrow spectrum linewidth was found in an STO consisting of a perpendicularly magnetized free layer and an in-plane magnetized reference layer \cite{kubota13} 
due to a high tunnel magnetoresistance (TMR) ratio \cite{yuasa04JJAP,parkin04,yuasa04} 
and large relative angle between the magnetizations in the free and reference layers. 
This type of STO will be a practical structure for the further development of spintronics devices based on STO, 
such as microwave generator and neuromorphic computing \cite{locatelli14,grollier16,torrejon17,kudo17,romera18,furuta18,tsunegi18,markovic19,tsunegi19}. 
However, the response of this type of STO to an oscillating signal has not been fully understood yet. 
It should be noted that it is not obvious whether this type of STO can be synchronized to an oscillating signal and/or show chaos.


In this paper, a theoretical study is comprehensively developed on the synchronization and chaos driven by alternating current or microwave field 
in an STO consisting of a perpendicularly magnetized free layer and an in-plane magnetized reference layer. 
The forced synchronization is observed in numerical simulation for wide ranges of amplitude and frequency of the alternative current. 
An analytical theory clarified that the locking range, as well as the phase difference between the STO and the current, is predominantly determined by the nonlinear frequency shift of the STO. 
It is also clarified that the spin-transfer torque asymmetry also plays a key role in widening the locking range. 
For an alternative current outside the synchronized region, chaos is found by evaluating the Lyapunov exponent. 
The microwave field is also found to cause the forced synchronization and chaos in the present STO. 
It should, however, be noted that the microwave field should rotate in the same direction with the precession direction of the magnetization.
The parameter regions corresponding to chaos by the microwave field is found to be narrower than that by the alternating current. 


The paper is organized as follows. 
In Sec. \ref{sec:System description}, we give a brief description of the STO studied in this work. 
The definitions of the alternating current and microwave field are also given. 
In Sec. \ref{sec:Response to alternating current}, we investigate the response of the STO to the alternating current. 
In particular, the phase synchronization and chaos are studied both numerically and analytically.
Similarly, in Sec. \ref{sec:Response to microwave field}, we investigate the response of the STO to the microwave field. 
The summary of this work is in Sec. \ref{sec:Conclusion}.


\section{System description}
\label{sec:System description}


\begin{figure}[tb]
\centering
\includegraphics[width=5cm]{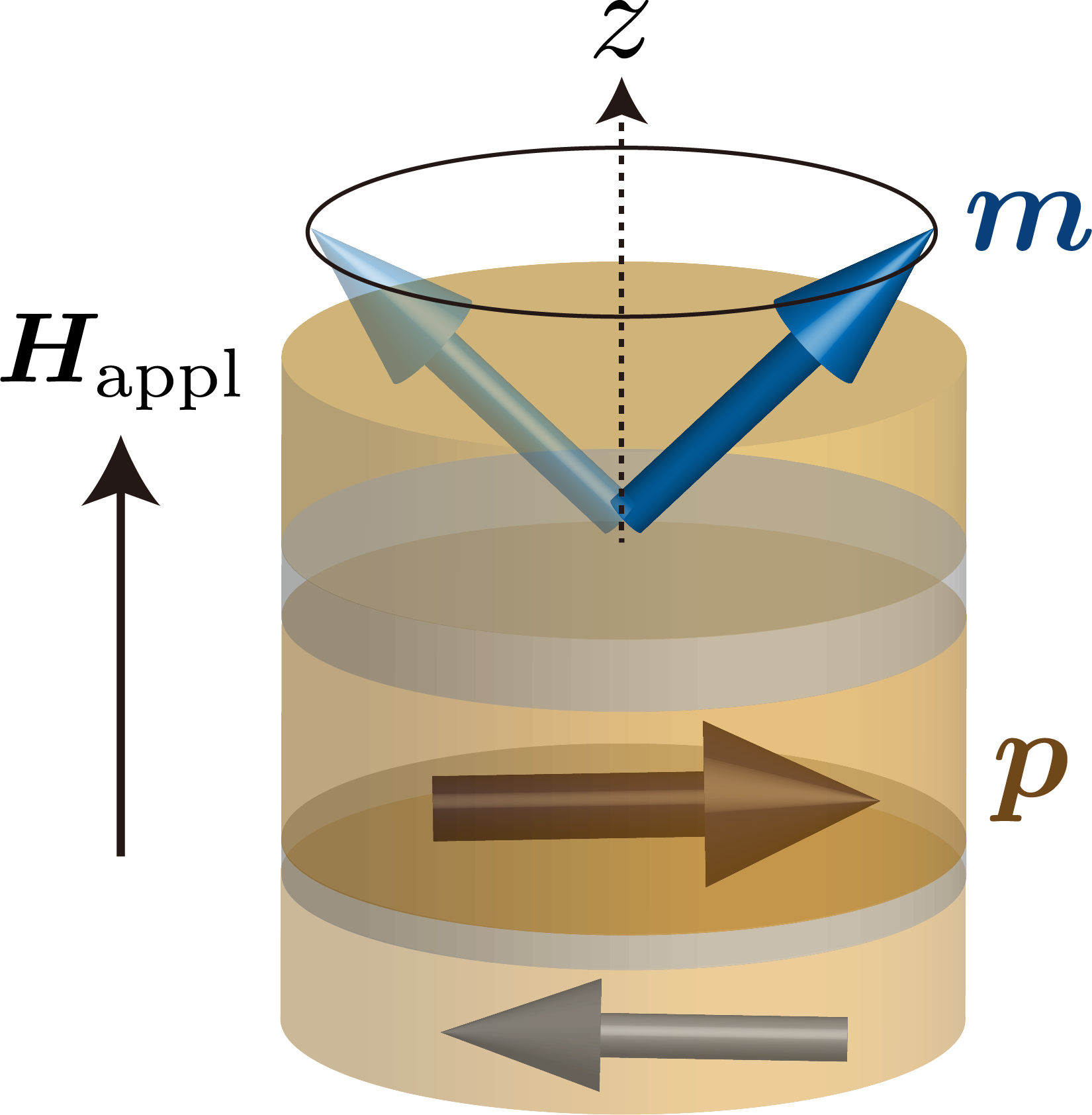}
\caption{
  Schematic view of the system. 
  The unit vectors pointing in the directions of the magnetizations in the free and reference layers are denoted as $\bm{m}$ and $\bm{p}$. 
  The magnetization $\bm{m}$ shows an auto-oscillation around the $z$ axis, 
  whereas $\bm{p}$ is parallel to the $x$ axis. 
  An external field $H_{\rm appl}$ is applied to the $z$ direction. 
  The bottom layer below the reference layer is the pinned layer, where its magnetization is antiferromagnetically coupled to that in the reference layer. 
}
\label{fig:fig1}
\end{figure}


The system we are interested in is an STO consisting of a perpendicularly magnetized free layer 
and an in-plane magnetized reference layer \cite{kubota13}, schematically shown in Fig. \ref{fig:fig1}. 
We denote the unit vector pointing in the magnetization direction in the free layer as $\bm{m}$, 
where the $z$ axis is perpendicular to the film plane, whereas the $x$ axis is parallel to the magnetization $\bm{p}$ ($|\bm{p}|=1$) in the reference layer. 
The electric current $j$ flows along the $z$ direction, 
where the positive current corresponds to the electrons flowing from the free to reference layer. 
The magnetization dynamics in the free layer is described by the Landau-Lifshitz-Gilbert (LLG) equation given by 
\begin{equation}
\begin{split}
  \frac{d \bm{m}}{d t}
  =&
  - \gamma 
  \bm{m} 
  \times \bm{H}
  - 
  \frac{\gamma \hbar g(\bm{m},\bm{p}) j}{2 e M V} 
  \bm{m} 
  \times 
  \left(
    \bm{p} 
    \times 
    \bm{m}
  \right)
\\
  &
  + 
  \alpha \bm{m} 
  \times 
  \frac{d \bm{m}}{d t},
  \label{eq:LLG}
\end{split}
\end{equation}
where $\gamma$ and $\alpha$ are the gyromagnetic ratio and the Gilbert damping constant, respectively. 
The saturation magnetization and volume of the free layer are denoted as $M$ and $V$, respectively. 
The magnetic field $\bm{H}$ consists of an external field $H_{\rm appl}$ and the magnetic anisotropy field along the $z$-direction as
\begin{equation}
  \bm{H}
  =
  \begin{pmatrix}
    0 \\ 
    0 \\ 
    H_{\rm appl} + \left( H_{\rm K} - 4 \pi M \right) m_{z}
  \end{pmatrix},
  \label{eq:H}
\end{equation}
where $H_{\rm K}$ is the interfacial anisotropy field \cite{yakata09,ikeda10,kubota12} whereas $4\pi M$ is the shape anisotropy field. 
We neglect the stray field from the reference layer due to the following reason. 
In the experiment in Ref. \cite{kubota13}, the reference layer is placed onto a pinned layer. 
They are coupled antiferromagnetically due to the interlayer exchange coupling, as schematically shown in Fig. \ref{fig:fig1}. 
Thus, the stray field from the reference layer is approximately canceled by that from the pinned layer. 
As a result, the total stray field acts on the free layer becomes negligibly small compared with the perpendicular field. 
In addition, due to the presence of the pinned layer, 
we also assume that the magnetization direction of the reference layer is fixed. 
The factor $g(\bm{m},\bm{p})$ in the spin-transfer torque term in Eq. (\ref{eq:LLG}) is given by \cite{slonczewski05} 
\begin{equation}
  g (\bm{m},\bm{p}) 
  = 
  \frac{\eta}{1 + \lambda \bm{m} \cdot \bm{p}},
  \label{eq:g}
\end{equation}
where $\eta \ (0 < \eta < 1)$ is the spin polarization of the current 
whereas $\lambda \ (0 < \lambda < 1)$ characterizes the angular dependence of the spin-transfer torque. 
The form of Eq. (\ref{eq:g}) was found in both the giant magnetoresistive (GMR) system and magnetic tunnel junction (MTJ) \cite{slonczewski96,slonczewski05,slonczewski89,brataas01,stiles02JAP,kovalev02,xiao04,theodonis06}.
The parameter $\lambda$ in the GMR system characterizes the difference of the spin accumulation, 
as well as the spin-dependent transmission probability, between the parallel (P) and antiparallel (AP) alignments, 
of the magnetizations in the free and reference layer \cite{brataas01,stiles02JAP,kovalev02,xiao04}, 
whereas it in MTJ is related to the density of state of the ferromagnet and 
determines the difference of the tunnel conductance at the P and AP alignments \cite{slonczewski05,slonczewski89,theodonis06}. 
Because of this factor, the magnetization switching current from AP to P alignment by the spin-transfer effect is smaller than that from P to AP alignment \cite{katine00,kubota05}. 
In this paper, we call $\lambda$ the spin-transfer torque asymmetry 
because a finite $\lambda$ makes the magnitudes of the spin-transfer torque different for a certain $m_{x}$ and $-m_{x}$. 
According to Ref. \cite{slonczewski05}, we use $\lambda=\eta^{2}$. 
It is useful for the readers to note that a finite $\lambda$ is necessary to excite a limit cycle oscillation in the STO \cite{taniguchi13}. 
The values of the parameters used in this work are derived from the experiment in Ref. \cite{kubota13}, as well as its theoretical analysis \cite{taniguchi13,taniguchi17}, as 
$\gamma = 1.764 \times 10^{7}$ rad/(Oe s), 
$M=1448.3$ emu/c.c., 
$H_{\rm appl} = 2 \times 10^{3}$ Oe, 
$H_{\rm K} = 1.8616 \times 10^{4}$ Oe, 
$\eta = 0.537$, $\lambda = 0.288$, 
$V = \pi \times 60 \times 60 \times 2$ nm${}^{3}$, and $\alpha=0.005$. 
In the following sections, Eq. (\ref{eq:LLG}) will be solved both numerically and analytically. 


In this work, we study the response of the STO to an alternating current and microwave field. 
In the presence of the alternating current, the total current $j$ is given by 
\begin{equation}
  j 
  = 
  j_{\rm dc} 
  + 
  j_{\rm ac} 
  \cos \Omega t, 
  \label{eq:ACcurrent}
\end{equation}
where $j_{\rm dc}$ is the direct current, 
whereas $j_{\rm ac}$ and $f_{\rm ac}=\Omega/(2\pi)$ are the amplitude and frequency of the alternating current. 
On the other hand, in the presence of the microwave field, the field $\bm{H}_{\rm ac}$, 
\begin{equation}
  \bm{H}_{\rm ac}
  =
  \begin{pmatrix}
    H_{\rm ac}^{x} \cos \Omega t \\
    H_{\rm ac}^{y} \cos (\Omega t + \phi_{\rm ac})	\\
    0
  \end{pmatrix},
  \label{eq:ACfield}
\end{equation}
should be added to the magnetic field $\bm{H}$. 
The amplitudes of the microwave field in the $x$ and $y$ directions are denoted as $H_{\rm ac}^{x}$ and $H_{\rm ac}^{y}$, respectively, 
whereas the phase difference between them is $\phi_{\rm ac}$. 
We note that superposition of two microwave fields with the phase difference as such is realized experimentally 
by using two coplanar wave guides \cite{suto17}. 
In Sec. \ref{sec:Response to microwave field}, we will investigate four cases, 
i.e., the linear polarized microwave field along the $x$ direction($H_{\rm ac}^{x}=H_{\rm ac}$ and $H_{\rm ac}^{y}=0$), 
that along the $y$ direction ($H_{\rm ac}^{x}=0$ and $H_{\rm ac}^{y}=H_{\rm ac}$), 
the circularly polarized microwave field rotating in the counterclockwise direction with respect to the positive $z$ direction ($H_{\rm ac}^{x}=H_{\rm ac}^{y}=H_{\rm ac}$ and $\phi_{\rm ac}=-90^{\circ}$), 
and that rotating in the clockwise direction ($H_{\rm ac}^{x}=H_{\rm ac}^{y}=H_{\rm ac}$ and $\phi_{\rm ac}=90^{\circ}$). 


The electric current injected into an STO excites not only the spin-transfer torque, called Slonczewski torque, in Eq. (\ref{eq:LLG}) 
but also the other torque called fieldlike torque, 
which also originates from the spin-transfer from the conducting electrons to the magnetization.
The fieldlike torque acts as a torque due to the in-plane magnetic field. 
Throughout this work, however, we neglect the fieldlike torque due to the following reasons.
First, the effective field of the fieldlike torque due to the direct current is negligibly smaller than the perpendicular field.
This is because we focus on the auto-oscillation state of the magnetization.
The auto-oscillation is excited when the Slonczewski torque balances with the damping torque.
Since the damping constant $\alpha$ is small, 
the magnitude of the Slonczewski torque is much smaller than the perpendicular field.
In addition, the fieldlike torque is at least one order of magnitude smaller than the Slonczewski torque.
Regarding these facts, the fieldlike torque in the present system is negligible.
Second, the fieldlike torque due to the alternating current might play a role on synchronization even if its magnitude is small.
In fact, as shown below, the forced synchronization occurs even for a small alternating current.
It is, however, useful to note that the role of the fieldlike torque due to the alternating current is identical 
to that of the linearly polarized microwave field along the $x$ direction.
Since the forced synchronization due to the linearly polarized microwave field will be extensively studied 
in Sec. \ref{sec:Response to microwave field},
the fieldlike torque due to the alternating current is not discussed in this paper.


\section{Response to alternating current}
\label{sec:Response to alternating current}

In this section, we investigate the response of the STO to an alternating current. 


\subsection{Numerical simulation of phase synchronization}

\begin{figure}[tb]
\centering
\includegraphics[width=8.5cm]{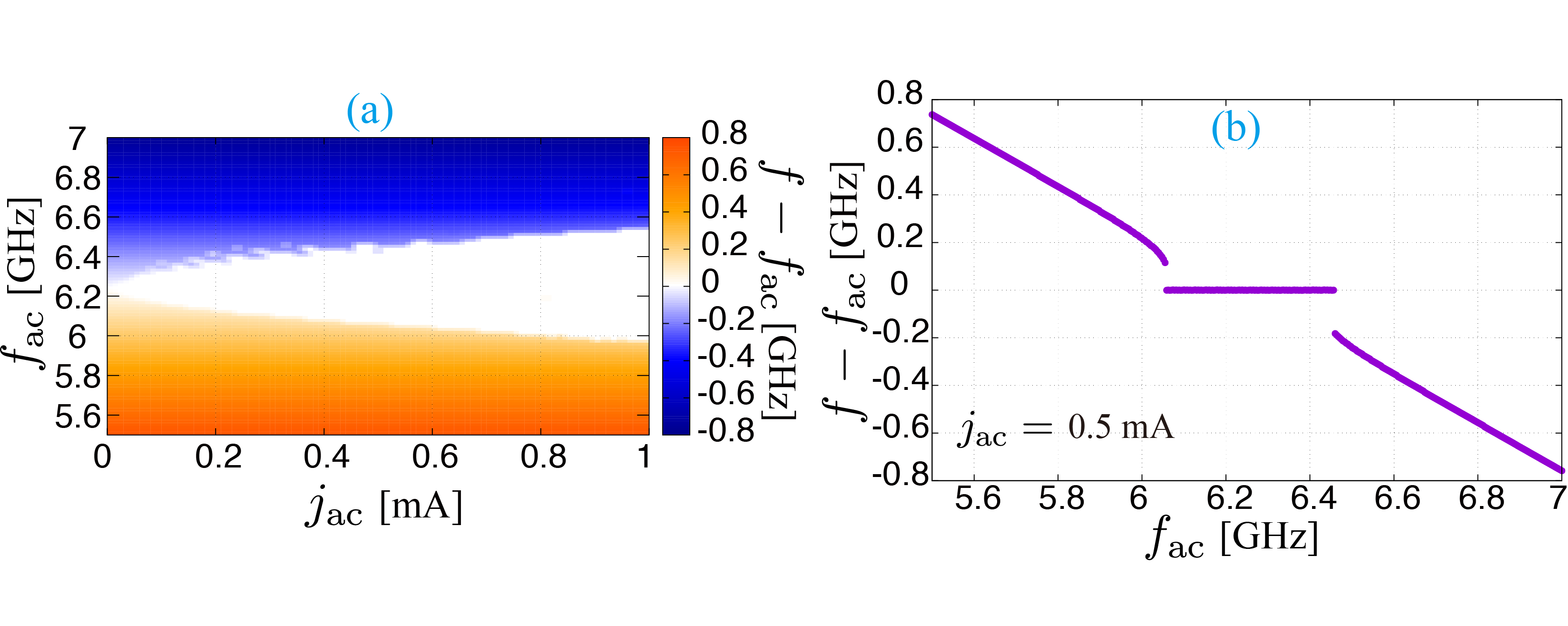}
\caption{
  (a) Dependence of the frequency detuning $f-f_{\rm ac}$ on the amplitude ($j_{\rm ac}$) and frequency ($f_{\rm ac}$) of the alternating current, 
  where $f$ is the oscillation frequency of the STO defined as the peak frequency of the Fourier spectrum of $m_{x}(t)$ obtained from numerical simulation. 
      The direct current is fixed to $j_{\rm dc}=2.5$ mA, corresponding to the free-running frequency of $6.2$ GHz. 
      The zero detuning corresponds to the synchronization of the STO to the alternating current. 
  (b) The frequency detuning as a function of $f_{\rm ac}$ at $j_{\rm ac}=0.5$ mA. 
}
\label{fig:fig2}
\end{figure}

Here we show the results of the numerical simulation of Eq. (\ref{eq:LLG}). 
Figure \ref{fig:fig2}(a) shows the dependencies of 
the frequency detuning between the frequencies of STO ($f$) and alternating current, $f-f_{\rm ac}$, 
on the magnitude ($j_{\rm ac}$) and frequency ($f_{\rm ac}$) of the alternating current.
Here the frequency $f$ of the STO in the numerical simulation is defined as the maximum peak frequency of the Fourier spectrum of $m_{x}(t)$.
The direct current is fixed to $j_{\rm dc}=2.5$ mA, corresponding to the free-running frequency of $6.2$ GHz. 
The zero detuning region corresponds to the forced synchronization of the STO to the alternating current. 
It can be seen from Fig. \ref{fig:fig2}(a) that the locking range increases nearly linear with the magnitude of the alternating current. 
A sawtooth structure appears near the boundary of the locking region, in particular in a higher frequency side. 
Such a structure has been found in previous works focusing on other types of STO \cite{finocchio12,bortolotti13,iacocca13}, 
however the origin is not fully explained yet. 
Figure \ref{fig:fig2}(b) shows the frequency detuning as a function of the current frequency, where $j_{\rm ac}=0.5$ mA. 
The locking range is nearly $0.4$ GHz, which is comparable to that found in a different type of STO \cite{georges08}. 


\begin{figure}[tbh]
\centering
\includegraphics[width=8.5cm]{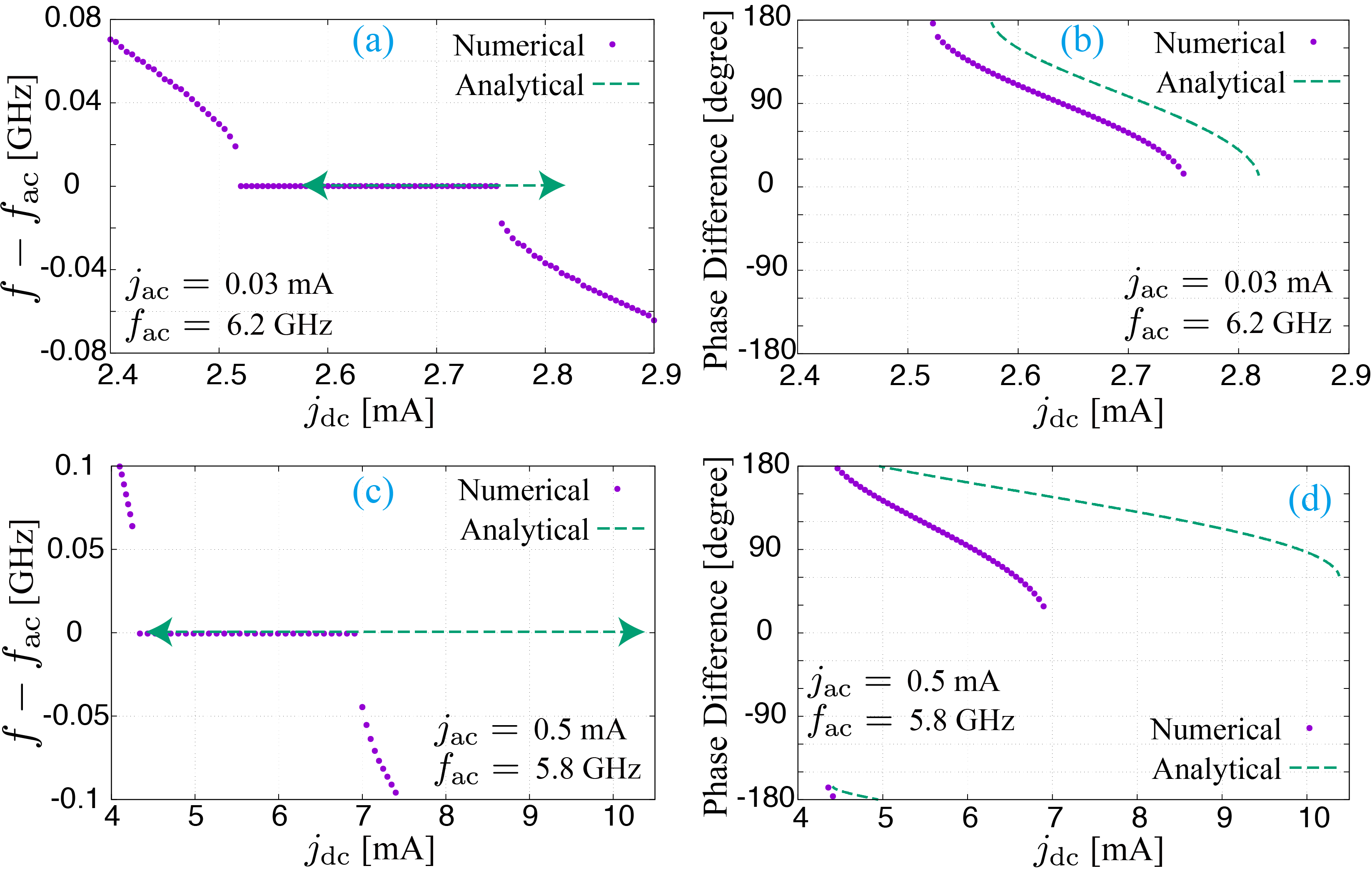}
\caption{
  (a) Frequency detuning and (b) phase difference between the STO and the alternating current as a function of the direct current. 
      The amplitude and frequency of the alternating current are fixed to $j_{\rm ac}=0.03$ mA and $f_{\rm ac}=6.2$ GHz. 
	  The purple dots are obtained from the numerical simulation, 
	  whereas the green dashed lines correspond to the estimations by an analytical theory. 
  The values of $j_{\rm ac}$ and $f_{\rm ac}$ are changed to $0.5$ mA and $5.8$ GHz in (c) and (d). 
}
\label{fig:fig3}
\end{figure}


\begin{figure}[tbh]
\centering
\includegraphics[width=7cm]{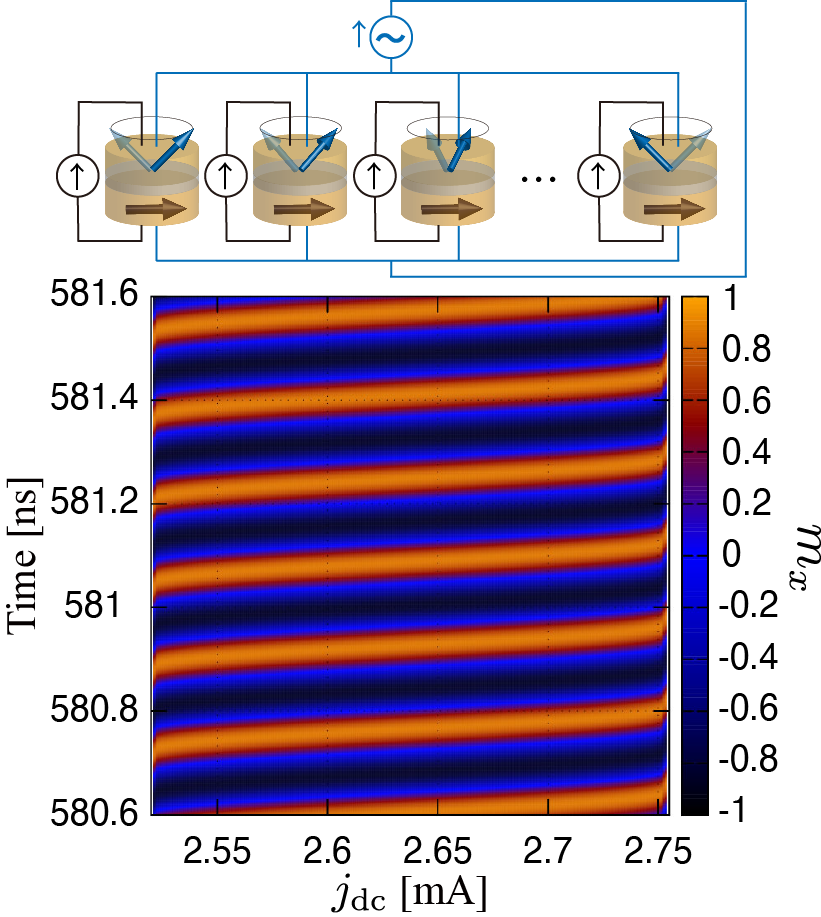}
\caption{
  Schematic picture of an array of STOs is shown above, where the pinned layers are not shown, for simplicity. 
  The magnitude of the direct current injected into each STO is different. 
  Thus, when the oscillators are driven solely by the direct currents, the oscillation frequencies are different. 
  On the other hand, when a common alternating current is injected and the locking condition is satisfied, 
  all of the STOs oscillate with the identical frequency. 
  The phase of each STO, however, depends on the magnitude of the direct current. 
  Time evolution of $m_{x}$ for various values of the direct current $j_{\rm dc}$ is shown below. 
  The same color corresponds to an isophase surface of the oscllating $m_{x}$. 
}
\label{fig:fig4}
\end{figure}


The above result indicates that the injection of the alternating current results in the frequency locking of the STO. 
We should note that the frequency locking can be measured easily in experiments and therefore has been extensively studied. 
However, the injection of the time-dependent signal to a nonlinear oscillator results not only in the frequency synchronization but also in phase synchronization \cite{rippard05,pikovsky03}. 
The control of the oscillator's phase is becoming important because of the development of the practical devices utilizing it, 
such as phased array radar \cite{sun13} and neuromorphic devices \cite{kudo17,markovic19,tsunegi19}. 
Therefore, we are also inclined to investigate the phase of the STO in the presence of the alternating current. 


Figure \ref{fig:fig3}(a) and \ref{fig:fig3}(b) show the frequency detuning and phase difference as a function of the direct current. 
The amplitude and frequency of the alternating current are fixed to $j_{\rm ac}=0.03$ mA and $f_{\rm ac}=6.2$ GHz, respectively. 
Figure \ref{fig:fig3}(a) indicates that the frequency synchronization occurs in the current range of $2.53 \lesssim j_{\rm dc} \lesssim 2.75$ mA. 
Note that the phase difference in Fig. \ref{fig:fig3}(b) is defined only in the synchronized state. 
It is shown that the phase difference varies over wide range from $0^{\circ}$ to $180^{\circ}$. 
At the center of the locking region, the phase difference becomes as close as $90^{\circ}$. 
In the next section, it will be clarified that the value $90^{\circ}$ at the center is related to the frequency nonlinearity of the STO. 


The frequency detuning and phase difference at different values of $j_{\rm ac}$ and $f_{\rm ac}$ are shown in Figs. \ref{fig:fig3}(c) and \ref{fig:fig3}(d), respectively, 
where $j_{\rm ac}=0.5$ mA and $f_{\rm ac}=5.8$ GHz. 
As already revealed in Fig. \ref{fig:fig2}, the locking range increases due to the large $j_{\rm ac}$. 
The phase difference can be manipulated between $30^{\circ}$ and $180^{\circ}$, as well as $-180^{\circ}$ and $-150^{\circ}$. 
Note that the ranges of the direct current in Figs. \ref{fig:fig3}(b) and \ref{fig:fig3}(d) are different. 
This is because the free-running frequency of the STO changes by changing the magnitude of the direct current \cite{kubota13,taniguchi13,taniguchi17}, 
and the free-running frequency should be close to $f_{\rm ac}$ to excite synchronization. 
As will be clarified in the next section, an analytical theory of nonlinear oscillator well explains 
the phase synchronization for a relatively high $f_{\rm ac}$ shown in Fig. \ref{fig:fig3}(b), 
whereas that for a relatively low $f_{\rm ac}$ shown in Fig. \ref{fig:fig3}(d) has a deviation from the analytical theory. 


The results shown in Fig. \ref{fig:fig3} indicate that the present STO can manipulate the phase of input signal, 
which is of great interest from both fundamental and applied physics. 
For example, the phased array radar \cite{sun13} is an electronic device manipulating a orientation of input wave signal. 
Figure \ref{fig:fig4} shows an example of such a refraction. 
Let us imagine that we have an array of the STO and apply different direct currents to the oscillators. 
Each STO then shows a limit cycle oscillation having different frequency. 
The injection of common alternating currents to such an STO array leads to the phase locking. 
Importantly, the output signal from each STO will have a different phase with respect to the alternative current, 
as can be seen in Figs. \ref{fig:fig3}(b) and  \ref{fig:fig3}(d). 
As a result, the isophase surface of the output signals from the STOs points to an oblique direction, 
as can be seen in Fig. \ref{fig:fig4}, 
indicating that the array of the STOs results in the refraction of the input signal. 
As can be seen in this example, STO acts as a phase manipulator. 


\subsection{Analytical theory of phase synchronization}

To clarify the role of the alternating current on the phase locking, 
let us develop an analytical theory of the synchronization. 
We introduce new variables, $N$ and $\varphi$, as $N=1-m_{z}$ and $\varphi=\tan^{-1}(m_{y}/m_{x})$. 
The relation between $m_{z}$ and $N$ is similar to the Holstein-Primakoff transformation \cite{holstein40} 
between the spin operator and bosonic creation-annihilation operator, 
whereas $\varphi$ is the phase of the in-plane components. 
The LLG equation, Eq. (\ref{eq:LLG}), can be expressed in terms of $N$ and $\varphi$ as 
\begin{equation}
\begin{split}
  \dot{N}
  =&
  - \frac{\gamma \hbar \eta \sqrt{N(2-N)}(1-N) j \cos \varphi}{2 eMV (1 + \lambda \sqrt{N(2-N)} \cos \varphi)}
\\
  & -
  \alpha N (2-N) \dot{\varphi},
  \label{eq:Ampeq}
\end{split}
\end{equation}
\begin{equation}
\begin{split}
  \dot{\varphi}
  =&
  \omega(N)
\\
  &+ 
  \frac{\gamma \hbar \eta j \sin \varphi}{2 eMV \sqrt{N(2-N)} (1+\lambda \sqrt{N(2-N)} \cos \varphi)}.
  \label{eq:phaseeq}
\end{split}
\end{equation}
Note that $\omega(N)=\gamma [H_{\rm appl}+(H_{\rm K}-4\pi M)(1-N)]$ is the angular velocity of the magnetization in a conservative system. 
In the absence of the alternating current, the magnetization shows a limit cycle oscillation 
when the direct current $j_{\rm dc}$ is larger than a critical value, $j_{\rm c}$, as clarified in Ref. \cite{taniguchi13}. 
The amplitude $N$ in the oscillation state shows a small oscillation around an averaged value \cite{taniguchi17}, 
which is denoted as $n_{0}$ through this paper. 
The current necessary to excite a limit cycle oscillation with the averaged amplitude $n_{0}$ is given by \cite{taniguchi13} (see also Appendix \ref{sec:AppendixA})) 
\begin{equation}
  j_{\rm dc}
  =
  - \frac{2eMV \alpha \lambda}{\gamma \hbar \eta}
  \frac{n_{0} \omega_{0} (2-n_{0}) \sqrt{1-\lambda^{2}n_{0}(2-n_{0})}}
    {(1-n_{0})(\sqrt{1-\lambda^{2}n_{0}(2-n_{0})}-1)}, 
  \label{eq:current_n0}
\end{equation}
whereas the oscillation frequency for $n_{0}$ is given by 
$f_{0}=\omega_{0}/(2\pi)=\gamma \left[ H_{\rm appl} + (H_{\rm K} - 4 \pi M) (1-n_{0}) \right]/(2\pi)$.


On the other hand, in the presence of the alternating current, the amplitude $N$ shows a deviation $\delta n$ from the averaged value $n_{0}$. 
In this case, the equations of motion for $\delta n$ and $\varphi$ up to the first order of $\delta n$ are given by 
\begin{equation}
\begin{split}
  \dot{\delta n}
  =&
  - (C_{s} + C_{\alpha}) 
  \delta n
\\
  & -
  \frac{\gamma \hbar \eta \sqrt{n_{0}(2-n_{0})}(1-n_{0}) j_{\rm ac} \cos \varphi \cos \Omega t}
    {2eMV (1+ \lambda\sqrt{n_{0}(2-n_{0})} \cos \varphi)},
  \label{eq:deltan}
\end{split}
\end{equation}
\begin{equation}
\begin{split}
  \dot{\varphi}
  =&
  \omega_{0} 
  - 
  \omega_{\rm K} 
  \delta n
\\
  & +
  \frac{\gamma \hbar \eta j_{\rm ac} \sin \varphi \cos \Omega t}
    {2eMV \sqrt{n_{0}(2-n_{0})}(1+\lambda\sqrt{n_{0}(2-n_{0})} \cos \varphi)}.
  \label{eq:phase_deltan}
\end{split}
\end{equation}
Here, we introduce $\omega_{\rm K} = \gamma (H_{\rm K} - 4 \pi M)$ and 
\begin{equation}
\begin{split}
  C_{s}
  =&
  \frac{A j_{\rm dc}}{\lambda}
  \left[ 
    - 1 
    + 
    \frac{1}{\sqrt{1 - \lambda^{2}n_{0}(2-n_{0})}} 
  \right.
\\
  & 
  \left.
    - \lambda^{2} 
    \frac{(1-n_{0})^{2}}{ \left\{ 1 - \lambda^{2} n_{0}(2-n_{0}) \right\}^{3/2}}
  \right],
  \label{eq:Cs}
\end{split}
\end{equation}
\begin{equation}
\begin{split}
  C_{\alpha}
  =&
  2 \alpha (1 - n_{0}) \omega_{0} - \alpha n_{0} (2 - n_{0}) \omega_{\rm K},
  \label{eq:Calpha}
\end{split}
\end{equation}
where $A=\gamma\hbar\eta/(2eMV)$. 
Since we are interested in the synchronized state, we assume that $\varphi=\Omega t+\phi$, where $\phi$ is the phase difference between the STO and the alternating current. 
Note that $\delta n$ should satisfy $\delta n(t)=\delta n(t+T_{\rm ac})$ because of the periodicity of the synchronization, 
where $T_{\rm ac}=1/f_{\rm ac}$ is the period of the alternating current. 
Therefore, averaging Eqs. (\ref{eq:deltan}) and (\ref{eq:phase_deltan}) over the period of $T_{\rm ac}$ 
and substituting Eq. (\ref{eq:deltan}) into (\ref{eq:phase_deltan}), we obtain 
\begin{equation}
\begin{split}
  \Delta
  =
  \frac{-Aj_{\rm ac} \left[ Z(1+\Lambda)\cos\phi+(1-\Lambda)\sin\phi \right]}
    {2 \sqrt{n_{0}(2-n_{0})[1-\lambda^{2}n_{0}(2-n_{0})]}},
  \label{eq:Delta_equation}
\end{split}
\end{equation}
where $\Delta=\omega_{0}-\Omega$ and 
\begin{equation}
  Z = 
  \frac{\omega_{\rm K} n_{0} (2-n_{0}) (1-n_{0})}{C_{s} + C_{\alpha}},
  \label{eq:Z}
\end{equation}
\begin{equation}
  \Lambda 
  = 
  \frac{(1 - \sqrt{1 - \lambda^{2}n_{0}(2-n_{0})})^{2}}{\lambda^{2}n_{0}(2-n_{0})}
  \label{eq:Lambda}
\end{equation}
Equation (\ref{eq:Delta_equation}) indicates that the forced synchronization to the alternating current occurs 
when the frequency detuning is in the range of 
\begin{equation}
  |\Delta|
  \leq
  \frac{\gamma \hbar \eta j_{\rm ac} (1+\Lambda) \sqrt{Z^{2} + r^{2}}}
    {4 eMV \sqrt{n_{0}(2-n_{0})(1 - \lambda^{2} n_{0}(2-n_{0}))}},
  \label{eq:Delta_leq}
\end{equation}
with the phase difference given by 
\begin{equation}
\begin{split}
  \phi
  =&
  - \phi_{0}
  + 
  \sin^{-1}
  \left[ 
    - \Delta
    \frac{4 eMV}{\gamma \hbar \eta j_{\rm ac}}
  \right.
\\
  & 
  \times
  \left.
    \frac{\sqrt{n_{0}(2-n_{0})(1-\lambda^{2}n_{0}(2-n_{0}))}}
      {(1+ \Lambda) \sqrt{Z^{2} + r^{2}}}
    \right].
  \label{eq:Phi_Aopen}
\end{split}
\end{equation}
Here, we introduce 
\begin{equation}
  r 
  = 
  \frac{1-\Lambda}{1+\Lambda},
  \label{eq:r}
\end{equation}
whereas $\phi_{0}$ satisfies 
\begin{align}
  \sin \phi_{0} 
  = 
  \frac{Z}{\sqrt{Z^{2} + r^{2}}}, 
&&
  \cos \phi_{0} 
  = 
  \frac{r}{\sqrt{Z^{2} + r^{2}}}.
  \label{eq:phi0}
\end{align}
Since $0< \Lambda < 1$ from its definition, we note that $0 < r < 1$. 


We emphasize that $\phi_{0}$ is close to $90^{\circ}$. 
This is because of the relation $Z \gg 1 > r$. 
In fact, for example, near the equilibrium state at $n_{0}=0$, $Z$ is estimated to be 
\begin{equation}
  \lim_{n_{0}\to 0}
  Z
  =
  \frac{\omega_{\rm K}}{\alpha \left[ (1-3\lambda^{2}/2)\omega_{0}' - \omega_{\rm K} \right]}
  \simeq
  48.9,
  \label{eq:Z_crit}
\end{equation}
where $\omega_{0}^{\prime} = \gamma (H_{\rm appl} + H_{\rm K} - 4 \pi M)$. 
The parameter $Z$ is a key quantity of the forced synchronization 
because it determines the locking range, as can be seen in Eq. (\ref{eq:Delta_leq}). 
The parameter $Z$ is called dimensionless nonlinear frequency shift \cite{slavin09}. 
In fact, $Z$ depends on $\omega_{\rm K}$, which determines the frequency shift by the change of the oscillation amplitude. 
We emphasize, however, that the parameter $Z$ in the present STO depends not only on the nonlinear frequency shift $\omega_{\rm K}$ but also on other parameters. 
In particular, Eq. (\ref{eq:Z_crit}) reveals that the spin-transfer torque asymmetry, $\lambda$, determines the bandwidth. 
We emphasize that the role of the spin-transfer torque asymmetry $\lambda$ on the synchronization is firstly clarified in this paper. 


We study the applicability of the analytical theory by comparing it with the numerical simulation shown in the last section. 
It should be noted that the phase difference at which the free-running frequency of the STO is identical to $f_{\rm ac}$ is close to $90^{\circ}$, 
see Figs. \ref{fig:fig3}(b) and \ref{fig:fig3}(d). 
Note that the analytical estimation from Eq. (\ref{eq:Phi_Aopen}) is $\lim_{\Delta \to 0}\phi \simeq 90^{\circ}$ (see Appendix \ref{sec:AppendixB}). 
At this point, the analytical theory well explains the numerical simulation. 
In Fig. \ref{fig:fig3}, we have shown the analytical estimations of the locking range and phase difference. 
As can be seen in Figs. \ref{fig:fig3}(a) and \ref{fig:fig3}(b), the analytical theory 
well explains the locking range and the phase difference for a relatively small alternating current and high frequency. 
On the other hand, the theory shows deviations for a relatively large alternating current and low frequency, as in Figs. \ref{fig:fig3}(c) and \ref{fig:fig3}(d). 
There are two reasons for the deviations between the numerical simulation and the analytical theory. 
First, we assume that the deviation $\delta n$ of the amplitude $N$ from its averaged value $n_{0}$ is small in the derivation of Eq. (\ref{eq:Delta_leq}). 
However, for a large alternating current, higher order terms of $\delta n$ are not negligible. 
Second, we also assume that the free-running frequency of the STO in the numerical simulation is identical to the analytical estimation, $f_{0}$. 
However, as clarified in Ref. \cite{taniguchi17}, the free-running frequency of the STO differs from $f_{0}$, especially in a low frequency region, 
or equivalently in a large amplitude region, 
although the analytical estimation of the averaged amplitude $n_{0}$ still works in this region. 
This difference originates from the fact that the spin-transfer torque has a projection to the precession direction when the oscillation amplitude becomes large. 
Due to these reasons, the analytical theory does not work for a large alternating current and/or low frequency regions. 


\subsection{Chaos}

In this section, we investigate chaos caused by the alternating current. 
The chaos in this paper is defined as a dynamics with a positive Lyapunov exponent. 
The evaluation method of the (maximum) Lyapunov exponent is summarized in Appendix \ref{sec:AppendixC}. 


\begin{figure}[tb]
\centering
\includegraphics[width=8.5cm]{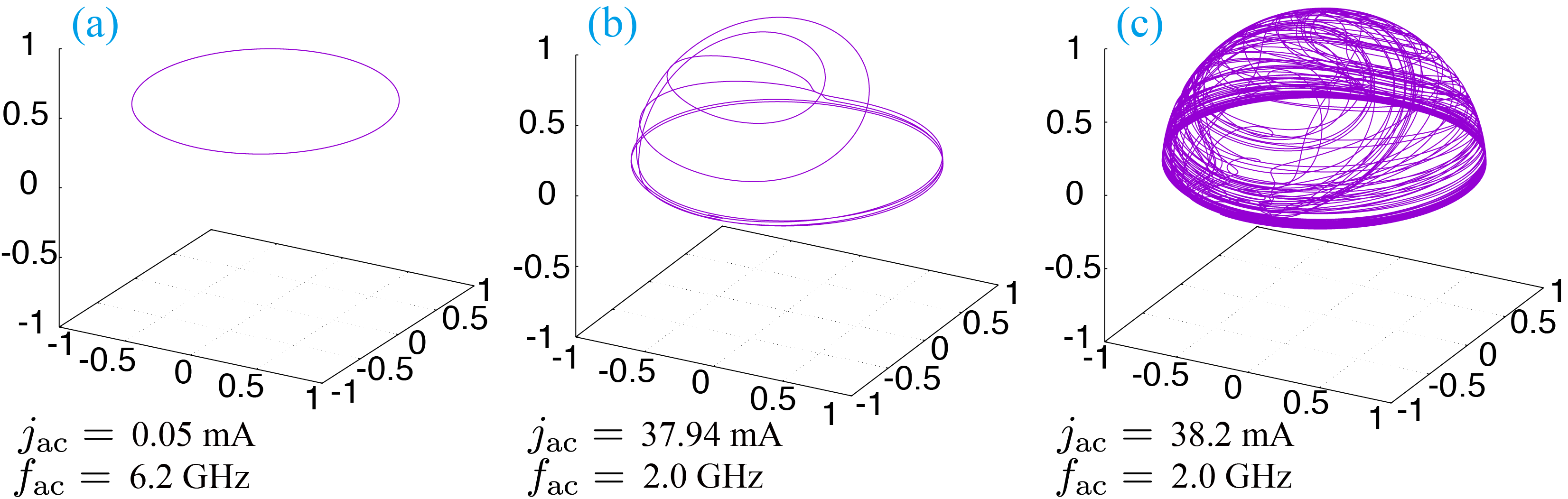}
\caption{
         The trajectories of magnetization in
	 (a) a limit cycle oscillation ($j_{\rm ac} = 0.05$ mA, $f_{\rm ac} = 6.2$ GHz),
	 (b) an amplitude modulation on limit cycle oscillation ($j_{\rm ac} = 37.94$ mA, $f_{\rm ac} = 3.0$ GHz),and 
	 (c) a chaos state ($j_{\rm ac} = 38.2$ mA, $f_{\rm ac} = 2.0$ GHz).
}
\label{fig:fig5}
\end{figure}



\begin{figure}[tb]
\centering
\includegraphics[width=8.5cm]{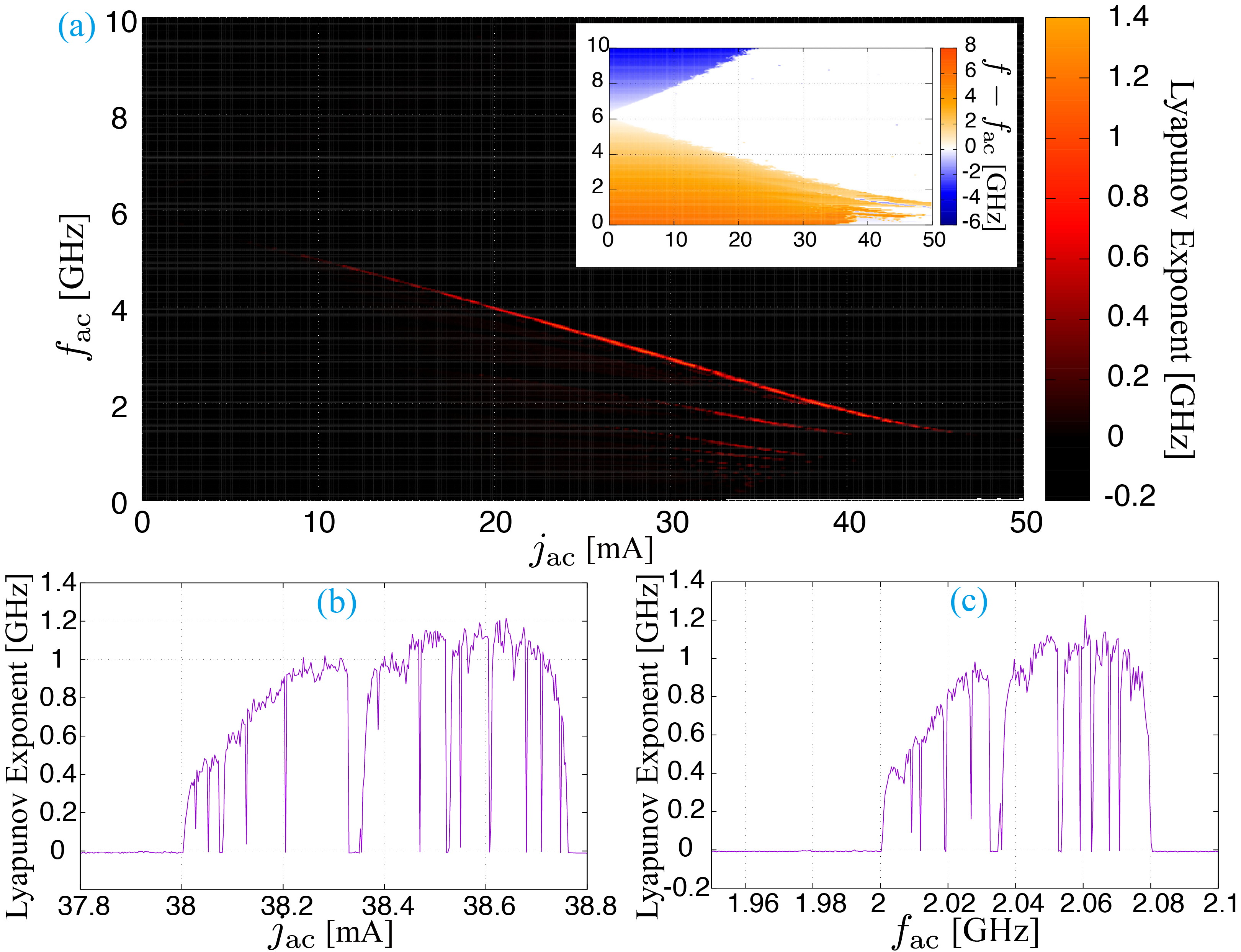}
\caption{
  (a) The Lyapunov exponent as functions of $j_{\rm ac}$ and $f_{\rm ac}$. 
      The inset shows the frequency detuning in the same $j_{\rm ac}$ and $f_{\rm ac}$ regions. 
  The Lyapunov exponent at $f_{\rm ac}=2.0$ GHz and $j_{\rm ac}=38$ mA are shown in (b) and (c), respectively. 
}
\label{fig:fig6}
\end{figure}


Figures \ref{fig:fig5}(a)-\ref{fig:fig5}(c) show typical dynamical trajectories of the magnetization in the presence of an alternating current.
A limit cycle oscillation with a unique amplitude shown in Fig. \ref{fig:fig5}(a) was extensively studied in the previous sections.
With increasing the amplitude of the alternating current, a periodic oscillation with amplitude modulation appears, as shown in Fig. \ref{fig:fig5}(b).
In addition, a nonperiodic complex dynamics shown in Fig. \ref{fig:fig5}(c) also appears, depending on the amplitude and frequency of the alternating current.
As discussed below, these dynamics are distinguished by the Lyapunov exponent and bifurcation diagram.


Figure \ref{fig:fig6}(a) shows the Lyapunov exponent as functions of the magnitude ($j_{\rm ac}$) and frequency ($f_{\rm ac}$) of the alternating current, 
where the direct current is fixed to $j_{\rm dc}=2.5$ mA. 
Positive Lyapunov exponents are found in a large-alternating-current region, indicating the existence of chaos in the present STO. 
The inset in Fig. \ref{fig:fig6}(a) shows the frequency detuning in the same $j_{\rm ac}$ and $f_{\rm ac}$ regions. 
Comparing Fig. \ref{fig:fig6}(a) and its inset, we notice that chaos appears at the low-frequency region 
close to the boundary of the forced synchronization. 
Figures \ref{fig:fig6}(b) and \ref{fig:fig6}(c) show 
the dependencies of the Lyapunov exponent on $j_{\rm ac}$ and $f_{\rm ac}$, respectively, 
where $f_{\rm ac}=2.0$ GHz in Fig. \ref{fig:fig6}(b), whereas $j_{\rm ac}=38$ mA in \ref{fig:fig6}(c). 
It should be noted that the Lyapunov exponent is on the same order of magnitude of the free-running frequency of the STO. 
We remind the readers that the inverse of the Lyapunov exponent is the typical time 
at which two dynamical trajectories having slightly different initial conditions show a visible deviation \cite{strogatz01}. 
The fact that the Lyapunov exponent is on the same order of oscillation frequency of the STO indicates that 
the time necessary to show such a deviation is on the same order of the oscillation period. 
Therefore, a limit cycle oscillation is no longer stable in the chaos region, and the STO immediately shows a chaotic behavior 
when the alternating current is injected. 


\begin{figure}[tbh]
\centering
\includegraphics[width=8.5cm]{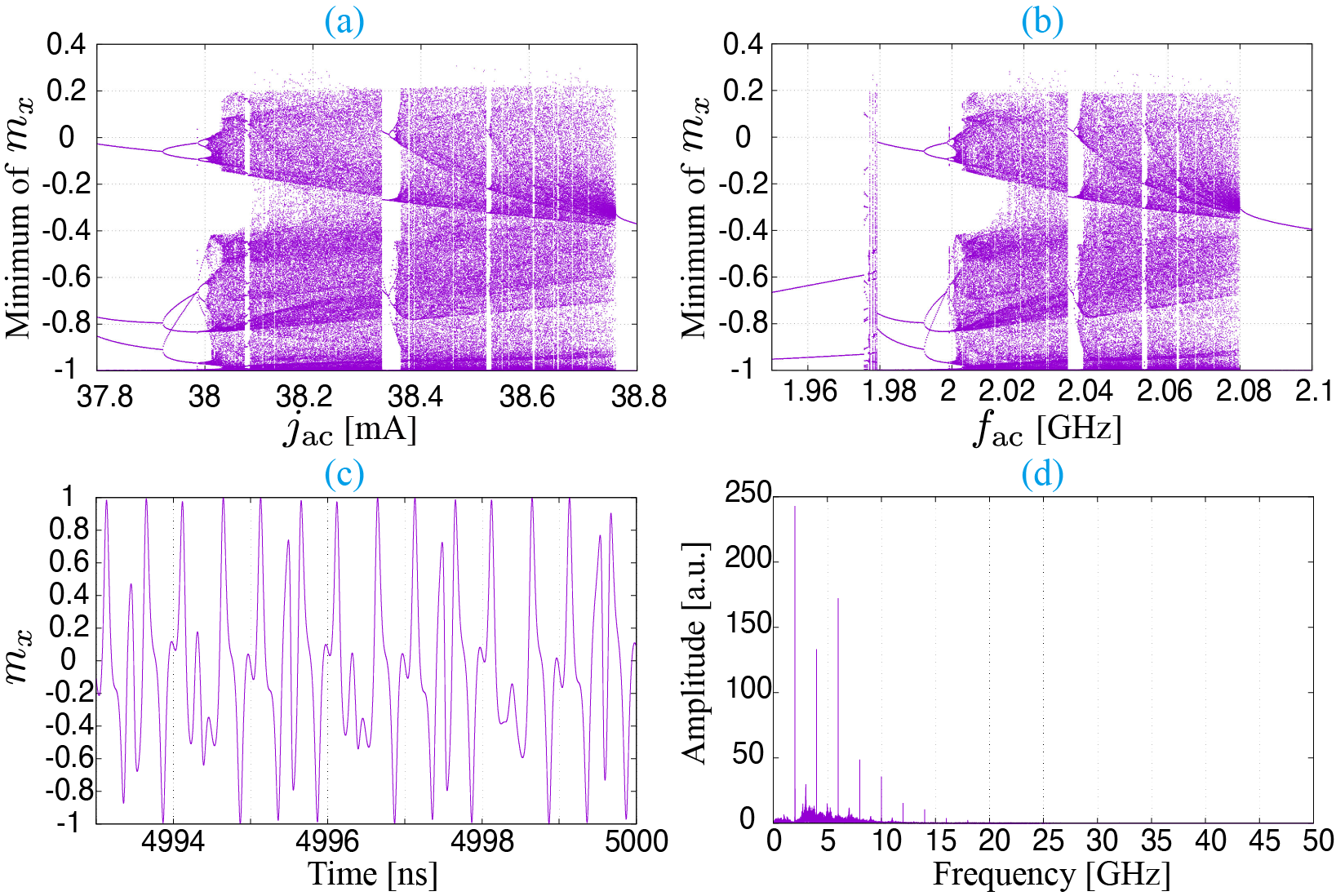}
\caption{
  The bifurcation diagrams for (a) $f_{\rm ac}=2.0$ GHz and (b) $j_{\rm ac}=38$ mA, 
  where temporal local minimum of $m_{x}$ are shown. 
  The time evolution of $m_{x}$ and its Fourier spectrum at $f_{\rm ac}=2.0$ GHz and $j_{\rm ac}=38.2$ mA 
  are shown in (c) and (d), respectively. 
}
\label{fig:fig7}
\end{figure}


In order to apprehend the dynamical trajectory in chaos, we evaluate $m_{x}$ at which local minimum temporarily exists. 
Figures \ref{fig:fig7}(a) and \ref{fig:fig7}(b) are the bifurcation diagrams of the local minimum of $m_{x}$ 
as functions of $j_{\rm ac}$ and $f_{\rm ac}$, respectively, 
where $f_{\rm ac}=2.0$ GHz in Fig. \ref{fig:fig7}(a) and $j_{\rm ac}=38$ mA in Fig. \ref{fig:fig7}(b). 
Comparing these figures with Figs. \ref{fig:fig6}(b) and \ref{fig:fig6}(c), 
we first notice that the local minimum of $m_{x}$ has multiple values even if the Lyapunov exponent is zero, and therefore, the dynamics is nonchaotic. 
This is because the alternating current outside the locking region provides an amplitude modulation on a limit cycle oscillation \cite{li06}. 
Second, the local minimum of $m_{x}$ shows a broad distribution when the Lyapunov exponent is positive, which is a typical structure in chaos \cite{strogatz01}. 
An example of the time evolution of $m_{x}$ and its Fourier spectrum at a chaos state are shown in Fig. \ref{fig:fig7}(c) and \ref{fig:fig7}(d), respectively. 
As can be seen in these figures, the dynamics becomes nonperiodic, and the oscillation frequency can no longer be determined uniquely. 
The window structure found in Figs. \ref{fig:fig6}(a) and \ref{fig:fig6}(b), 
for example around the region of $j_{\rm ac}=38.4$ mA in Fig. \ref{fig:fig6}(a), is also a typical characteristic in chaos system, 
where the Lyapunov exponent becomes zero in a small range of the parameters \cite{ott02}. 


We notice that the magnitude of the alternating current necessary to induce chaos is one order of magnitude larger than that of the direct current. 
It is experimentally difficult to apply such a large alternating current to a magnetic tunnel junction because of possibility to induce electrostatic destruction. 
Therefore, a metallic system maybe suitable to detect chaos in the present STO. 
We should also notice that the previous works clarify that an in-plane magnetized STO shows chaos 
by an alternating current one order of magnitude smaller than the direct current \cite{li06,yang07,xu08}. 
We consider the reason for the excitation of chaos with such small alternating current originates from 
the existence of two oscillation modes in the in-plane magnetized STO, i.e., the oscillations around the in-plane easy axis and the out-of-plane hard axis. 
In fact, chaos found in these previous works is related to the transition between these two modes. 
On the other hand, the present STO has a single oscillation mode, i.e., the oscillation around the $z$ axis. 
Such a simple structure of the present STO might be the reason why a large alternating current is necessary to induce chaos. 


\subsection{Short summary of this section}

In this section, we investigated the response of the STO to the alternating current. 
The forced synchronization was found for wide ranges of the amplitude and frequency of the alternating current. 
It was shown that the phase difference between the STO and the alternating current became $90^{\circ}$ 
when the free-running frequency of the STO was identical to the frequency of the current. 
The analytical theory clarified that this value of the phase difference was due to the large nonlinear frequency shift of the STO. 
It was also clarified that the spin-transfer torque asymmetry also played a key role for the enhancement of the locking range. 
The applicability of the analytical theory was also studied. 
In addition, the evaluation of the Lyapunov exponent indicated the existence of chaos in the STO.



\section{Response to microwave field}
\label{sec:Response to microwave field}

In this section, we investigate the response of the STO to a microwave field. 
We note that the amplitude of the microwave field used in this section, typically on the order of $100$ Oe at maximum, 
is comparable to that of the alternating current ($\sim 1$ mA) studied in Sec. \ref{sec:Response to alternating current}, 
and therefore the results shown in this section can be directly compared with those in Sec. \ref{sec:Response to alternating current}; 
see Appendix \ref{sec:AppendixD}. 
In addition, as mentioned in Sec. \ref{sec:System description}, the fieldlike torque due to the alternating current 
plays the same role with the torque due to the linearly polarized microwave field in the $x$ direction. 
The estimation of the effective field of the fieldlike torque is also discussed in Appendix \ref{sec:AppendixD}.


\subsection{Numerical simulation of phase synchronization}

\begin{figure}[th]
\centering
\includegraphics[width=8.5cm]{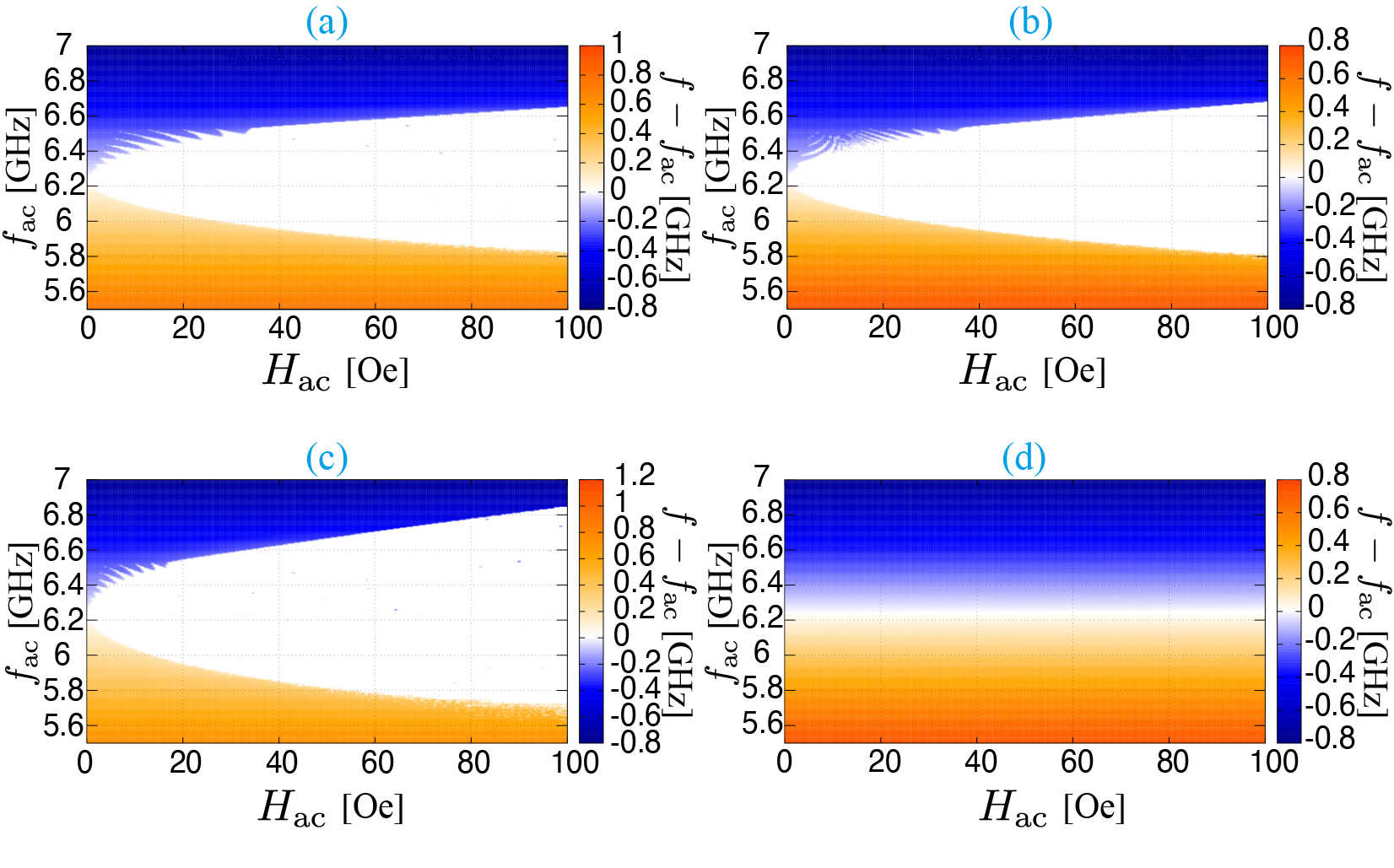}
\caption{
  Frequency detuning between the STO and microwave field as functions of the amplitude ($H_{\rm ac}$) and frequency ($f_{\rm ac}$) of the field. 
  The microwave fields are (a) linearly polarized in the $x$ direction, 
  (b) linearly polarized in the $y$ direction, 
  (c) circularly polarized rotating in the counterclockwise direction with respect to the positive $z$ direction, 
  and (d) circularly polarized rotating in the clockwise direction. 
  Note that the precession direction of the magnetization with respect to the positive $z$ direction is counterclockwise. 
  The direct current is fixed to $j_{\rm dc}=2.5$ mA. 
}
\label{fig:fig8}
\end{figure}

Let us show the results obtained by the numerical simulation of Eq. (\ref{eq:LLG}). 
Figures \ref{fig:fig8}(a)-\ref{fig:fig8}(d) show the dependencies of 
the frequency detuning between the STO and microwave field on the amplitude $H_{\rm ac}$ and the frequency $f_{\rm ac}$, 
where the microwave field are (a) linearly polarized in the $x$ direction, 
(b) linearly polarized in the $y$ direction, 
(c) circularly polarized rotating in the counterclockwise direction with respect to the positive $z$ direction, 
and (d) circularly polarized rotating in the clockwise direction. 
For the linearly polarized microwaves, the frequency locking occurs over wide ranges of $H_{\rm ac}$ and $f_{\rm ac}$. 
Similarly, the frequency locking is observed for the circularly polarized microwave field in the counterclockwise direction.
Since the circularly polarized microwave field is a superposition of two linearly polarized microwave fields, 
the locking range is larger than that of the linearly polarized field.
On the other hand, the locking does not occur for the microwave field rotating in the clockwise direction, as shown in Fig. \ref{fig:fig8}(d), 
where we note that the zero detuning in this figure appears only when the frequency of the microwave field is identical to the free-running frequency of the STO. 
The phenomenon as such in Fig. \ref{fig:fig8}(d) cannot be categorized as locking. 
It should be noted that the precession direction of the magnetization with respect to the positive $z$ direction is counterclockwise. 
In addition, the linearly polarized field is regarded as a superposition of the microwave fields rotating in the clockwise and counterclockwise directions. 
Therefore, the results shown in Fig. \ref{fig:fig8} indicates that 
the microwave field should have a component rotating in the same direction to the magnetization precession for the initiation of locking. 


\begin{figure}[t]
\centering
\includegraphics[width=8.5cm]{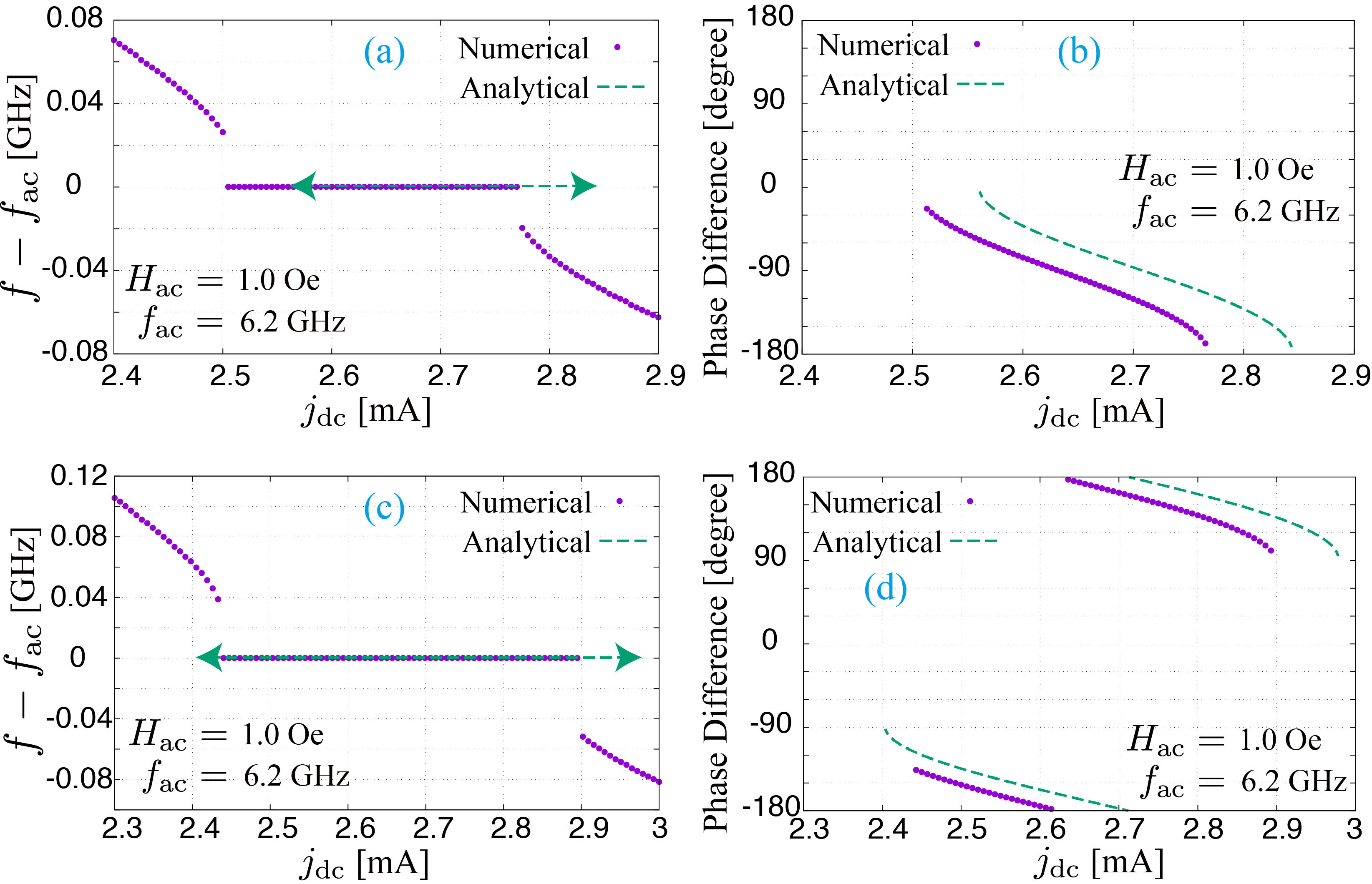}
\caption{
  (a) Frequency detuning and (b) phase difference as a function of the direct current for the linearly polarized microwave field in the $y$ direction. 
  (c) Frequency detuning and (d) phase difference as a function of the direct current for the circularly polarized microwave field 
  rotating in the counterclockwise direction with respect to the positive $z$ direction. 
  The amplitude and frequency of the microwave field are fixed to $H_{\rm ac}=1.0$ Oe and $f_{\rm ac}=6.2$ GHz. 
}
\label{fig:fig9}
\end{figure}


Figures \ref{fig:fig9}(a) and \ref{fig:fig9}(b) show the frequency detuning 
and the phase difference between the STO and the microwave field as a function of the direct current 
for the field linearly polarized in the $y$ direction. 
The amplitude and the frequency of the microwave field are fixed to $1.0$ Oe and $6.2$ GHz, respectively. 
For comparison, the frequency detuning and the phase difference for 
the circularly polarized microwave field rotating in the counterclockwise direction are shown in Figs. \ref{fig:fig9}(c) and \ref{fig:fig9}(d), respectively. 
The locking ranges for the linearly and circularly polarized microwaves are found to be nearly the same. 
On the other hand, the phase difference depends on the type of the microwave field. 
We also note that the dependence of the phase difference on the direct current differs from that found for the alternating current; see Fig. \ref{fig:fig3}(b). 
For example, the phase difference at the center of the locking range is $90^{\circ}$ for the alternating current, 
whereas it becomes $-90^{\circ}$ for the linearly polarized microwave field. 
In the next section, we develop a theory to clarify the role of the microwave field on the injection locking and the phase manipulation. 


\subsection{Analytical theory of phase synchronization}

Here, we develop an analytical theory of the phase synchronization induced by the microwave field. 
In the presence of the microwave field given by Eq. (\ref{eq:ACfield}), 
the equations of motion for $N$ and $\varphi$ are given by 
\begin{equation}
\begin{split}
  \dot{N}
  =&
  -\frac{ \gamma \hbar \eta \sqrt{N(2-N)} (1-N) j_{\rm dc} \cos \varphi }
    {2eMV (1 + \lambda \sqrt{N(2-N)} \cos \varphi)}
  - 
  \alpha 
  N (2-N) 
  \dot{\varphi}
\\
  &
  - \gamma H_{\rm ac}^{x} 
  \sqrt{N(2-N)} 
  \sin \varphi \cos \Omega t
\\
  & 
  + \gamma H_{\rm ac}^{y} 
  \sqrt{N(2-N)} 
  \cos \varphi \cos (\Omega t + \phi_{\rm ac}), 
  \label{eq:Amp_Hac}
\end{split}
\end{equation}
\begin{equation}
\begin{split}
  \dot{\varphi}
  =&
  \omega (N) 
  + 
  \frac{\gamma \hbar \eta  j_{\rm dc} \sin \varphi}{2 eMV\sqrt{N(2-N)} (1 + \lambda \sqrt{N(2-N)} \cos \varphi)}
\\
  &
  -\frac{1-N}{\sqrt{N(2-N)}}
  \left[ 
    \gamma H_{\rm ac}^{x} 
    \cos \varphi \cos \Omega t
  \right.
\\
  & 
  \left.
    + \gamma H_{\rm ac}^{y} 
    \sin \varphi \cos (\Omega t + \phi_{\rm ac})
  \right]. 
  \label{eq:phase_Hac}
\end{split}
\end{equation}
As in the case of the response to the alternating current, 
we assume a small deviation $\delta n$ of the amplitude $N$ from the averaged value $n_{0}$, 
and reduce the equation of motion up to the first order of $\delta n$ as 
\begin{equation}
\begin{split}
  \delta 
  \dot{n}
  =&
  - \left(
    C_{s} + C_{\alpha}
  \right) 
  \delta n
  - 
  \gamma H_{\rm ac}^{x} 
  \sqrt{n_{0}(2-n_{0})} 
  \sin \varphi \cos \Omega t
\\
  &
  + \gamma H_{\rm ac}^{y} 
  \sqrt{n_{0}(2-n_{0})} 
  \sin \varphi \cos (\Omega t + \phi_{\rm ac}), 
  \label{eq:deltan_Hac}
\end{split}
\end{equation}
\begin{equation}
\begin{split}
  \dot{\varphi}
  =&
  \omega_{0} 
  - 
  \omega_{\rm K} 
  \delta n
  - 
  \frac{1-n_{0}}{\sqrt{n_{0}(2-n_{0})}}
\\
  & 
  \times
  \left[ 
    \gamma H_{\rm ac}^{x} 
    \cos \varphi \cos \Omega t
    + 
    \gamma H_{\rm ac}^{y} 
    \sin \varphi \cos (\Omega t + \phi_{\rm ac})
  \right].
  \label{eq:phase_deltan_Hac}
\end{split}
\end{equation}
Substituting Eq. (\ref{eq:deltan_Hac}) into Eq. (\ref{eq:phase_deltan_Hac}) and averaging over the period of $T_{\rm ac}$, 
we find that 
\begin{equation}
\begin{split}
  \Delta
  =&
  - \frac{\gamma (1-n_{0}) \sqrt{Z_{m}^{2} + 1}}{2 \sqrt{n_{0}(2-n_{0})}}
  \sin (\phi + \phi_{1} + \phi_{2})
\\
  & 
  \times
  \sqrt{(H_{\rm ac}^{x})^{2} + (H_{\rm ac}^{y})^{2} - 2 H_{\rm ac}^{x} H_{\rm ac}^{y} \sin \phi_{\rm ac}}, 
  \label{eq:Delta_phi_Hac}
\end{split}
\end{equation}
where we introduce 
\begin{equation}
\begin{split}
  Z_{m}
  &=
  \frac{Z}{(1-n_{0})^{2}}
\\
  &= 
  \frac{\omega_{\rm K} n_{0} (2-n_{0})}{(C_{s} + C_{\alpha}) (1-n_{0})}, 
  \label{eq:def_Zm}
\end{split}
\end{equation}
whereas $\phi_{1}$ and $\phi_{2}$ satisfy 
\begin{align}
  \cos \phi_{1} 
  = 
  \frac{Z_{m}}{\sqrt{Z_{m}^{2} + 1}},
&&
  \sin \phi_{1} 
  = 
  - \frac{1}{\sqrt{Z_{m}^{2} + 1}},
  \label{eq:def_phi1}
\end{align}
\begin{equation}
  \cos \phi_{2}
  =
  \frac{ H_{\rm ac}^{x} - H_{\rm ac}^{y} \sin \phi_{\rm ac} }
    {\sqrt{(H_{\rm ac}^{x})^{2} + (H_{\rm ac}^{y})^{2} - 2 H_{\rm ac}^{x} H_{\rm ac}^{y} \sin \phi_{\rm ac}}},
  \label{eq:def_phi2_cos}
\end{equation}
\begin{equation}
  \sin \phi_{2}
  =
  - \frac{ H_{\rm ac}^{y} \cos \phi_{\rm ac} }
    {\sqrt{(H_{\rm ac}^{x})^{2} + (H_{\rm ac}^{y})^{2} - 2 H_{\rm ac}^{x} H_{\rm ac}^{y} \sin \phi_{\rm ac}}}. 
  \label{eq:def_phi2_sin}
\end{equation}
We again emphasize that the parameter $Z_{m}$ plays key role in the determination of the locking range and the phase difference. 
As can be seen in Eq. (\ref{eq:def_Zm}), $Z_{m}$ is proportional to $Z$ given by Eq. (\ref{eq:Z}). 
Therefore, similarly to the case of the synchronization by the alternating current, 
the nonlinear frequency shift $\omega_{\rm K}$ and the spin-transfer torque asymmetry $\lambda$ are key parameters 
for widening the locking range and manipulating the STO phase. 


For the linearly polarized microwave field ($\phi_{\rm ac}=0^{\circ}$), 
Eq. (\ref{eq:Delta_phi_Hac}) becomes 
\begin{equation}
\begin{split}
  \Delta
  =&
  - \frac{\gamma (1-n_{0}) \sqrt{Z_{m}^{2} + 1}}{2 \sqrt{n_{0}(2-n_{0})}}
  \sin (\phi + \phi_{1} + \phi_{2})
\\
  & 
  \times
  \sqrt{(H_{\rm ac}^{x})^{2} + (H_{\rm ac}^{y})^{2}},
  \label{eq:Delta_phi_Hac_linear}
\end{split}
\end{equation}
where $\phi_{2}$ can be expressed as 
\begin{equation}
  \cos \phi_{2}
  =
  \frac{H_{\rm ac}^{x}}{\sqrt{(H_{\rm ac}^{x})^{2} + (H_{\rm ac}^{y})^{2}}},
  \label{eq:phi2_linear_cos}
\end{equation}
\begin{equation}
  \sin \phi_{2}
  =
  - \frac{H_{\rm ac}^{y}}{\sqrt{(H_{\rm ac}^{x})^{2} + (H_{\rm ac}^{y})^{2}}}.
  \label{eq:phi2_linear_sin}
\end{equation}
The injection locking occurs when 
\begin{equation}
  |\Delta| 
  \leq
  \frac{\gamma \sqrt{(H_{\rm ac}^{x})^{2} + (H_{\rm ac}^{y})^{2}}(1-n_{0}) \sqrt{Z_{m}^{2} + 1}}{2 \sqrt{n_{0}(2-n_{0})}}.
  \label{eq:lock_linear}
\end{equation}
In particular, for the linearly polarized microwave field along the $y$ direction ($H_{\rm ac}^{x}=0$ and $H_{\rm ac}^{y}=H_{\rm ac}$), 
$\phi_{2}= -90^{\circ}$. 
Therefore, the phase difference between the STO and the microwave field is 
\begin{equation}
  \phi
  = 
  \frac{\pi}{2}  
  - 
  \phi_{1}
  + 
  \sin^{-1} 
  \left[ 
    - \Delta 
    \frac{2 \sqrt{n_{0}(2-n_{0})}}{\gamma H_{\rm ac} (1-n_{0}) \sqrt{Z_{m}^{2} + 1}}
  \right].
  \label{eq:phi_linear}
\end{equation}
In addition, since $Z_{m} \gg 1$, $\phi_{1}\simeq 0$. 
As a result, the phase difference is expected to be close to $-90^{\circ}$ for $\Delta=0$ (see Appendix \ref{sec:AppendixB}). 
In Figs. \ref{fig:fig9}(a) and \ref{fig:fig9}(b), 
we also show the analytical estimation of the locking range and the phase difference. 
The result indicates that the analytical theory well explains the numerical simulation. 


On the other hand, for the circularly polarized microwave field in the counterclockwise direction, 
$H_{\rm ac}^{x}=H_{\rm ac}^{y}=H_{\rm ac}$ and $\phi_{\rm ac}=-90^{\circ}$. 
In this case, $\phi_{2}=0^{\circ}$, and Eq. (\ref{eq:Delta_phi_Hac}) becomes 
\begin{equation}
  \Delta
  =
  -\frac{\gamma H_{\rm ac} (1-n_{0}) \sqrt{Z_{m}^{2}+1}}{2 \sqrt{n_{0}(1-n_{0})}}
  \sin\phi, 
  \label{eq:phi_circular}
\end{equation}
where we use $\phi_{1} \simeq 0^{\circ}$. 
The comparisons between the numerical simulation and analytical estimation on the locking range and phase difference 
shown in Figs. \ref{fig:fig9}(c) and \ref{fig:fig9}(d) guarantees the applicability of 
the analytical theory developed above. 
The phase difference at $\Delta=0$ is $\pm 180^{\circ}$ (see also Appendix \ref{sec:AppendixB}). 
The difference of the phase difference at $\Delta=0$ between Figs. \ref{fig:fig9}(b) and \ref{fig:fig9}(d) comes from $\phi_{2}$. 


We also note that Eq. (\ref{eq:Delta_phi_Hac}) becomes $\Delta =0$ 
for a circularly polarized microwave field in the clockwise direction ($\phi_{\rm ac}=90^{\circ}$), 
indicating that the microwave does not lock the STO. 
The conclusion is consistent with the numerical simulation.



\begin{figure}[tb]
\centering
\includegraphics[width=8.5cm]{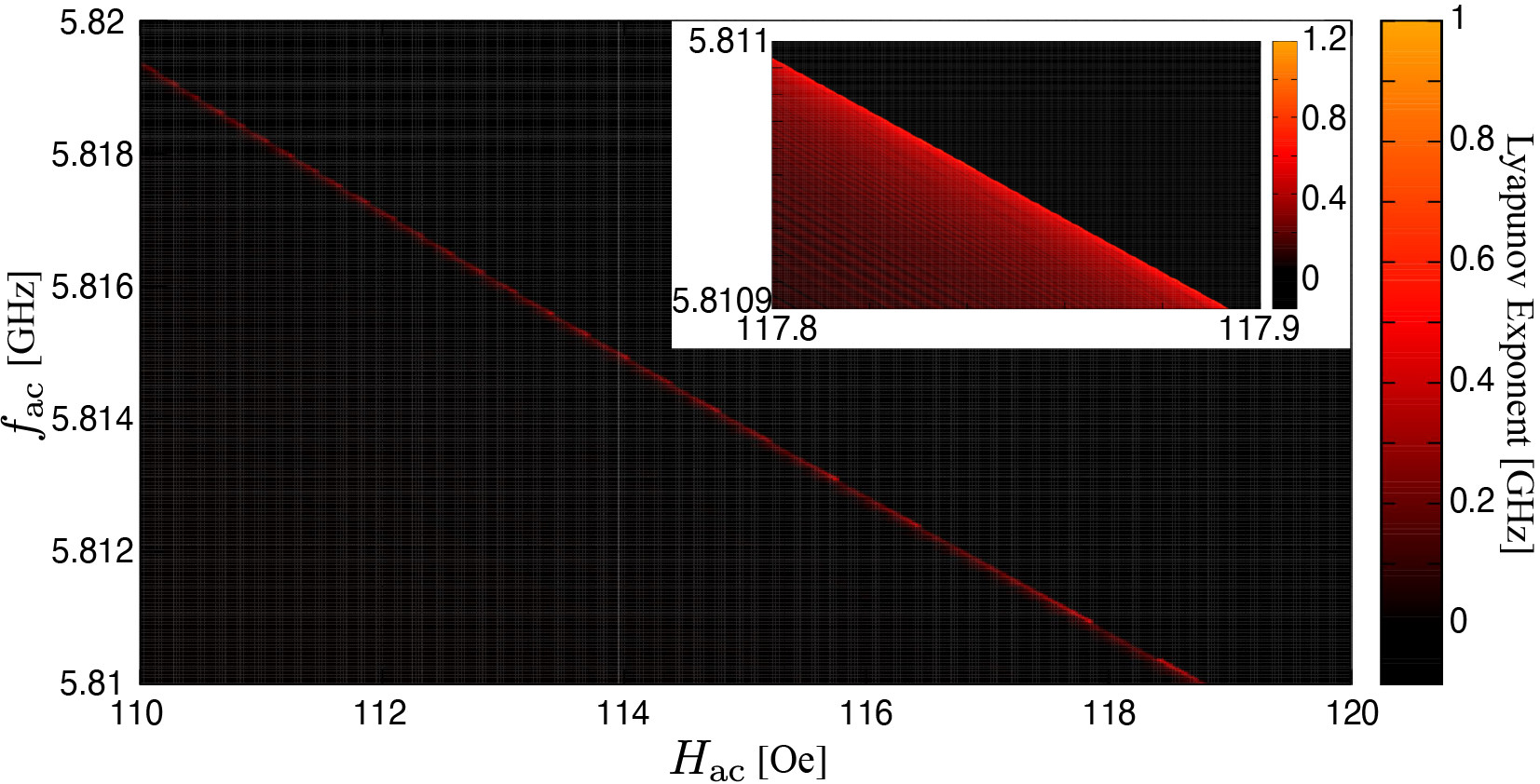}
\caption{
         The Lyapunov exponent as functions of $H_{\rm ac}$ and $f_{\rm ac}$. 
         The inset shows the Lyapunov exponent in narrow regions, $117.8 \le H_{\rm ac} \le 117.9$ Oe and $5.8019 \le f_{\rm ac} \le 5.811$ GHz. 
}
\label{fig:fig10}
\end{figure}


\subsection{Chaos}

\begin{figure}[tbh]
\centering
\includegraphics[width=8.5cm]{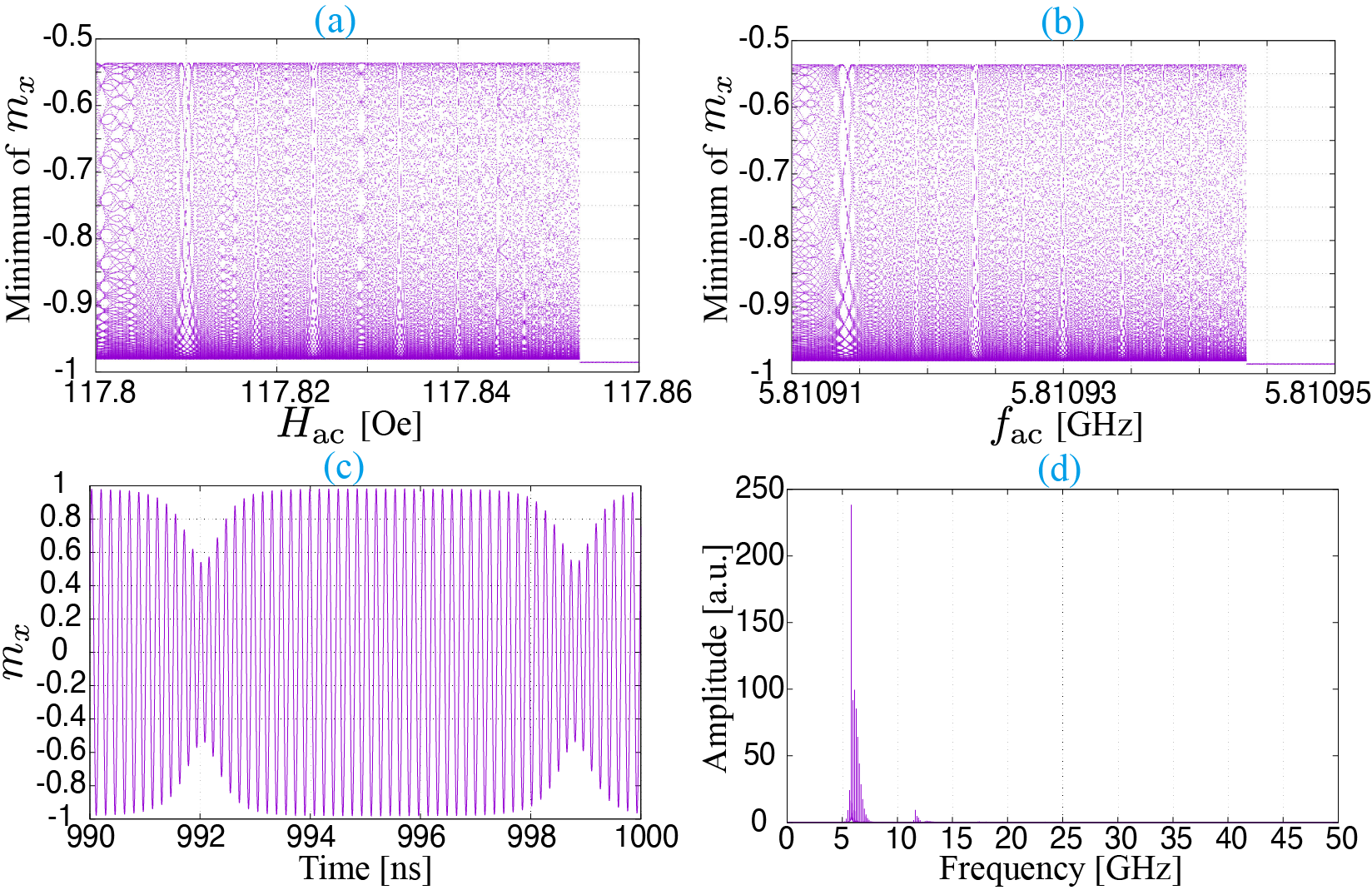}
\caption{
  The bifurcation diagrams for (a) $f_{\rm ac}=5.81094$ GHz and (b) $H_{\rm ac}=117.85$ Oe, 
  where temporal local minimum of $m_{x}$ are shown. 
  The time evolution of $m_{x}$ and its Fourier spectrum at $f_{\rm ac}=5.81094$ GHz and $H_{\rm ac}=117.85$ Oe are shown in (c) and (d), respectively. 
}
\label{fig:fig11}
\end{figure}

We investigate the possibility to cause chaos by the microwave field. 
Figure \ref{fig:fig10} shows the Lyapunov exponent as functions of the amplitude ($H_{\rm ac}$) and frequency ($f_{\rm ac}$) of the microwave field linearly polarized in $x$ direction. 
As in the case of the alternating current, a positive Lyapunov exponent is found in a lower frequency region outside the phase-locking range. 
However, the range corresponding to the positive Lyapunov exponent is narrower than that caused by the alternating current. 
For example, the chaos caused by the alternating current appears for $1 \lesssim f_{\rm ac} \lesssim 5$ GHz, 
i.e., chaos occurs for a frequency range of $4$ GHz; see Fig. \ref{fig:fig6}. 
On the other hand, the frequency range corresponding to chaos caused by the microwave field is only 0.01 GHz. 
Figures \ref{fig:fig11}(a) and \ref{fig:fig11}(b) are examples of the bifurcation diagrams 
as functions of the magnitude and frequency of the microwave field, respectively. 
Whereas the local minimum of $m_{x}$ shows multiple values, a structure typical in chaos, such as window, does not appear. 
Figures \ref{fig:fig11}(c) and \ref{fig:fig11}(d) are examples of the time evolution of $m_{x}$ and its Fourier spectrum, 
where the Lyapunov exponent is positive. 
Although the Fourier spectrum shows multiple peaks, the distribution is narrower than that caused by the alternating current shown in Fig. \ref{fig:fig7}(d). 
In view of these points, we consider that using the alternating current is more effective in inducing chaos in the present STO, compared with the use of the microwave field.


\subsection{Short summary of this section}

In this section, we investigated the response of the STO to the microwave field. 
It was shown both numerically and analytically that the forced synchronization occurs 
with exception for the case with circularly polarized microwave field rotating in the opposite direction to the precession of the magnetization. 
The analytical theory again revealed that the nonlinear frequency shift and the spin-transfer torque asymmetry determines the locking rage, 
as well as the phase difference between the STO and the microwave field. 
Although chaos was found, the parameter regions corresponding to chaos was narrower than that by the alternating current.


\section{Conclusion}
\label{sec:Conclusion}

In summary, the theoretical analysis was carried out for forced synchronization and chaos induced by an alternating current and microwave field 
in an STO consisting of a perpendicularly magnetized free layer and an in-plane magnetized reference layer. 
The alternating current induces phase synchronization over wide ranges of its amplitude and frequency, 
whereas the microwave field results in the synchronization 
when the field is linearly polarized or when the rotational direction of the circularly polarized field is identical to the precession direction of the magnetization. 
An analytical theory reveals that the nonlinear frequency shift, as well as the spin-transfer torque asymmetry, determines 
the locking range and phase difference between the oscillator and injected signal. 
The phase manipulation by the STO might open the way to new application such as phased array radar. 
Chaos was also found outside the synchronized region, where the frequency of the injected signal is lower than the free-running frequency of the oscillator. 
The positive Lyapunov exponent in chaos is on the same order of the free-running frequency, indicating that chaos is immediately initiated in the oscillator.


\section*{Acknowledgement}

This paper was based on the results obtained from a project (Innovative AI Chips and Next-Generation Computing Technology Development/(2) 
Development of next-generation computing technologies/Exploration of Neuromorphic Dynamics toward Future Symbiotic Society) commissioned by NEDO.



\appendix

\section{Derivation of Eq. (\ref{eq:current_n0})}
\label{sec:AppendixA} 

Here, we provide a brief description of the derivation of Eq. (\ref{eq:current_n0}). 
An auto-oscillation is excited in the STO when the energy injection by the Slonczewski torque balances with the dissipation due to the damping torque. 
In such a situation, the torque due to the magnetic field, $-\gamma \bm{m}\times\bm{H}$, becomes the dominant term in the LLG equation. 
The torque due to the field describes the oscillation of the magnetization on a constant energy surface. 
In the present system, the constant energy curve corresponds to the line with a constant $N$. 
Strictly speaking, $N$ is not constant in the auto-oscillation state because the Slonczewski and damping torques have different angular dependencies. 
However, since $\alpha$ is small, $N$ is approximately constant around the averaged value $n_{0}$, 
and, $dN/dt$ averaged over a precession period should be zero. 
Averaging Eq. (\ref{eq:Ampeq}) over $0 \le \varphi \le 2\pi$, we obtain 
\begin{equation}
\begin{split}
  \overline{
    \frac{dN}{dt}
  }
  =&
  \frac{-A(1-n_{0})j_{\rm dc}}{\lambda}
  \left(
    1
    -
    \frac{1}{\sqrt{1-\lambda^{2}n_{0}(2-n_{0})}}
  \right)
\\
  &-
  \alpha
  n_{0}
  \left(
    2
    -
    n_{0}
  \right)
  \omega_{0},
  \label{eq:current_n0_sub}
\end{split}
\end{equation}
where $\overline{\cdots}=\int_{0}^{2\pi}d\varphi/(2\pi)\cdots$, 
whereas $A=\gamma\hbar\eta/(2eMV)$ and $\omega_{0}=\gamma[H_{\rm appl}+(H_{\rm K}-4\pi M)(1-n_{0})]$. 
As mentioned above, $\overline{dN/dt}$ should be zero in the auto-oscillation state. 
Then, we obtain Eq. (\ref{eq:current_n0}) from Eq. (\ref{eq:current_n0_sub}), as done in Ref. \cite{taniguchi13}. 


\section{Analytical estimation of phase difference}
\label{sec:AppendixB}

As can be seen in Eqs. (\ref{eq:Phi_Aopen}), (\ref{eq:phi_linear}), and (\ref{eq:phi_circular}), 
the theoretical formulas of the phase include inverse trigonometric functions. 
The principal value of $\sin^{-1}$ in Eq. (\ref{eq:Phi_Aopen}) is defined in the range of $90^{\circ} \le \sin^{-1} \le 270^{\circ}$. 
In Eq. (\ref{eq:phi_linear}), on the other hand, 
the principal value is chosen to be $-270^{\circ} \le \sin^{-1} \le -90^{\circ}$, 
whereas in Eq. (\ref{eq:phi_circular}) it is in the range of $-180^{\circ} \le \sin^{-1} \le -90^{\circ}$ and $90^{\circ} \le \sin^{-1} \le 180^{\circ}$.


\section{Evaluation method of the Lyapunov exponent}
\label{sec:AppendixC}

This Appendix summarizes the evaluation method of the (maximum) Lyapunov exponent by the Shimada-Nagashima method \cite{shimada79} 
and its application to the LLG equation. 


\subsection{Shimada-Nagashima method}

A conventional definition of the Lyapunov exponent $\lambda_{\rm L}$ is 
$\delta x (t) = e^{\lambda_{\rm L}t} \delta x(0)$ \cite{strogatz01}, 
where $\delta x$ is a distance between two dynamic trajectories having slightly different initial conditions. 
This definition of the Lyapunov exponent is, however, not applicable directly to general nonlinear systems 
because the distance between two trajectories are, in general, not monotonically expanded. 
The Lyapunov exponent should be regarded as an average of the expansion rate over a certain timescale. 
Taking this into account, the Shimada-Nagashima method \cite{shimada79} evaluates the Lyapunov exponent 
as an average of an instantaneous expansion rate between the original trajectory and its perturbed one \cite{ott02,alligood97}. 


Let us define the Lyapunov exponent of a nonlinear dynamical system in $n$-dimensional phase space. 
We denote a dynamical trajectory obeying a nonlinear differential equation in this system as $\bm{x}(t)$. 
At a certain time $t_{0}$, we introduce a perturbed trajectory $\bm{x}^{\prime}(t_{0})$, 
which is obtained by moving $\bm{x}(t)$ to an arbitrary direction with $D[\bm{x}(t_{0}),\bm{x}^{\prime}(t_{0})]=\varepsilon$, 
where 
\begin{equation}
  D[\bm{x}(t), \bm{y}(t)]
  =
  \sqrt{
    \sum_{i}^{n}
    [x_{i}(t)-y_{i}(t)]^{2}
  },
  \label{eq:def_D}
\end{equation}
is the distance between two trajectories in the phase space. 
The parameter $\varepsilon$ is a small quantity. 
Two trajectories, $\bm{x}(t_{0})$ and $\bm{x}^{\prime}(t_{0})$, move to points $\bm{x}(t_{0}+\delta t)$ and $\bm{x}^{(1)}(t_{0}+\delta t)$ 
after a short time $\delta t$. 
Note that the distance between $\bm{x}(t_{0}+\delta t)$ and $\bm{x}^{(1)}(t_{0}+\delta t)$ is, in general, different from $\varepsilon$. 
Then, we define an instantaneous expansion rate as 
\begin{equation}
\begin{split}
  p_{1} 
  &= 
  \frac{D[\bm{x}(t_{0} + \delta t), \bm{x}^{(1)}(t_{0} + \delta t)]}{D[\bm{x}(t_{0}), \bm{x}^{(1)}(t_{0})]} 
\\
  &=
  \frac{D[\bm{x}(t_{0} + \delta t), \bm{x}^{(1)}(t_{0} + \delta t)]}{\varepsilon}. 
  \label{eq:def_p1}
\end{split}
\end{equation}
As mentioned above, the expansion rate depends on time. 
Therefore, we repeat the evaluation of the instantaneous expansion rate many times. 
At $t=t_{0}+\delta t$, we introduce 
\begin{align}
  \bm{x}^{(2)}(t_{0} + \delta t) 
  = 
  \bm{x}(t_{0} + \delta t) 
  + 
  \frac{\bm{x}^{(1)}(t_{0} + \delta t) - \bm{x}(t_{0} + \delta t)}{p_{1}}.
\end{align}
Note that $\bm{x}^{(2)}(t_{0}+\delta t)$ satisfies 
$D[\bm{x}(t_{0}+\delta t),\bm{x}^{(2)}(t_{0}+\delta t)]=\varepsilon$. 
The instantaneous expansion rate is then obtained as 
$p_{2}=D[\bm{x}(t_{0}+2 \delta t),\bm{x}^{(2)}(t_{0}+2\delta t)]/\varepsilon$. 
In general, the $k$th expansion rate $p_{k}$ is defined as 
\begin{equation}
  p_{k}
  =
  \frac{D[\bm{x}(t_{0}+k \delta t),\bm{x}^{(k)}(t_{0}+k \delta t)]}{\epsilon},
\end{equation}
where the $k$th perturbed variables $\bm{x}^{(k)}$ ($k>1$) obeys the recurrence formula,
\begin{equation}
\begin{split}
  \bm{x}^{(k+1)}(t_{0}+k \delta t)
  =&
  \bm{x}(t_{0}+k \delta t)
\\
  &+
  \frac{\bm{x}^{(k)}(t_{0}+k \delta t)-\bm{x}(t_{0}+k \delta t)}{p_{k}}.
  \label{eq:x_N_def}
\end{split}
\end{equation}
Note that $\bm{x}^{(k+1)}(t_{0}+k \delta t)$ satisfies 
$D[\bm{x}(t_{0}+k \delta t),\bm{x}^{(k+1)}(t_{0}+k \delta t)]=\varepsilon$. 
The (maximum) Lyapunov exponent is defined as the average of the instantaneous expansion rate as 
\begin{equation}
  \lambda_{\rm L}
  \equiv
  \lim_{k \to \infty} 
  \frac{1}{k \delta t}
  \sum_{i = 1}^{k} 
  \ln p_{i}.
  \label{eq:def_lambda}
\end{equation}


\subsection{Lyapunov exponent in nonautonomous system}

In the main text, we investigate the magnetization dynamics in the presence of a time-dependent signal. 
The Lyapunov exponent in a nonautonomous system as such kind is obtained by regarding the time in the time-dependent signal as a dynamical variable. 
We emphasize that this point is the key to identify chaos in STOs because of the following reason. 
It has been clarified that the limit cycle oscillation in an STO is well described by simplified models 
such as the macrospin model \cite{bertotti09text} and Thiele equation \cite{thiele73}. 
It should also be noted that magnetization dynamics in 
the other physical systems in spintronics and magnetism, such as a magnetic domain wall \cite{tatara04}, are also well described by a similar simplified model. 
These simplified models include two dynamical variables. 
For example, in the macrospin model, the number of the independent variable in $\bm{m}(t)$ is 2 
because, although $\bm{m}(t)$ is a vector in a three-dimensional system, the LLG equation conserves its magnitude, $|\bm{m}|=1$. 
The dynamical variables in the Thiele equation are the radius and phase of the vortex core, 
whereas those of the domain wall are represented by the position and magnetization tilted angle of the wall at its center. 
However, according to the Poincar\'e-Bendixson theorem \cite{strogatz01}, chaos is prohibited in a two-dimensional dynamical system. 
Therefore, an additional degree of freedom is necessary to excite chaos in these spintronics systems. 
The application of a time-dependent signal provides a solution to this issue because it makes the LLG equation nonautonomous 
and adds a new degree of freedom. 


The number of the dynamical variables in the present system becomes 3, $\bm{m}(t)$ and $\omega_{0}t$. 
We should note that the definitions of the distance and perturbation, given by Eqs. (\ref{eq:def_D}) and (\ref{eq:x_N_def}), cannot be applied 
in evaluating the Lyapunov exponent in the present system 
because, for example, the perturbation given by Eq. (\ref{eq:x_N_def}) does not conserves the norm of the magnetization $\bm{m}$, in general. 
Therefore, we define the distance between two trajectories as the angle between the magnetizations. 
In other words, we define the distance $\ell^{(1)}$ between $\bm{m}$ and its perturbation $\bm{m}^{(1)}$ at $t=t_{0}$ in a reduced phase space as 
\begin{equation}
  \ell^{(1)}[\bm{m}(t_{0}), \bm{m}^{(1)}(t_{0})]
  \equiv
  \cos^{-1} 
  \left[
    \bm{m}(t_{0}) 
    \cdot 
    \bm{m}^{(1)}(t_{0})
  \right].
  \label{eq:def_ellsp}
\end{equation}
The distance between two dynamical trajectories in the whole phase space is then given by 
\begin{equation}
\begin{split}
  &
  D[(\bm{m}(t_{0}),\omega_{0} t_{0}), (\bm{m}^{(1)}(t_{0}), \omega_{0} (t_{0}+\Delta t_{1})]
\\
  &=
  \sqrt{(\ell^{(1)}[\bm{m}(t_{0}), \bm{m}^{(1)}(t_{0})])^{2} + (\omega_{0} \Delta t_{1})^{2}}, 
  \label{eq:def_D_LLG}
\end{split}
\end{equation}
where $\Delta t_{1}$ is the perturbation to the time variable in the oscillating signal. 
As mentioned in the last section, the first perturbation at $t=t_{0}$ should satisfy
\begin{equation}
  D[(\bm{m}(t_{0}), \omega_{0} t_{0}), (\bm{m}^{(1)}(t_{0}), \omega_{0} (t_{0} + \Delta t_{1}))]
  =
  \varepsilon. 
  \label{eq:def_epsilon_LLG}
\end{equation}
The instantaneous expansion rate is then given by 
\begin{widetext}
\begin{equation}
\begin{split}
  p_{1}
  &=
  \frac{D[(\bm{m}(t_{0} + \delta t), \omega_{0} (t_{0} + \delta t)), (\bm{m}^{(1)}(t_{0} + \delta t), \omega_{0} (t_{0} + \delta t + \Delta t_{1}))]}{\varepsilon}
\\
  &=
  \frac{ \sqrt{ (\ell^{(1)\prime}[\bm{m}(t_{0}+\delta t),\bm{m}^{(1)}(t_{0}+\delta t)])^{2} + (\omega_{0}\Delta t_{1})^{2} }}{\varepsilon}, 
  \label{eq:def_p1_LLG}
\end{split}
\end{equation}
\end{widetext}
where $\ell^{(1)\prime}[\bm{m}(t_{0}+\delta t),\bm{m}^{(1)}(t_{0}+\delta t)]=\cos^{-1}[\bm{m}(t_{0}+\delta t)\cdot\bm{m}^{(1)}(t_{0}+\delta t)]$.

The second instantaneous expansion rate is obtained as follows. 
As mentioned in the last section, it is necessary to define $\bm{m}^{(2)}(t_{0}+\delta t)$ from $\bm{m}(t_{0}+\delta t)$ and $\bm{m}^{(1)}(t_{0}+\delta t)$. 
In addition, the perturbation $\Delta t_{2}$ to the time in the oscillating signal should also be defined. 
Note that $\bm{m}^{(2)}(t_{0}+\delta t)$ and $\Delta t_{2}$ should satisfy 
\begin{widetext}
\begin{align}
  D[(\bm{m}(t_{0}+\delta t), \omega_{0} (t_{0}+\delta t)), (\bm{m}^{(2)}(t_{0}+\delta t), \omega_{0} (t_{0} + \delta t + \Delta t_{2}))]
  =
  \varepsilon. 
\end{align}
\end{widetext}
Therefore, $\Delta t_{2}$ is given by 
\begin{equation}
  \omega_{0}
  \Delta t_{2}
  =
  \frac{\Delta t_{1}}{ p_{1} }.
\end{equation}
On the other hand, $\bm{m}^{(2)}(t_{0}+\delta t)$ is obtained by rotating $\bm{m}(t_{0}+\delta t)$ to 
the direction of $\bm{m}^{(1)}(t_{0}+\delta t)$ with the angle $\ell^{(2)}$, which is given by 
\begin{equation}
  \ell^{(2)}
  =
  \frac{\ell^{(1)\prime}}{p_{1}}.
\end{equation}

In general, the $k$-th expansion rate $p_{k}$ is defined as 
\begin{widetext}
\begin{equation}
\begin{split}
  p_{k}
  &=
  \frac{D[(\bm{m}(t_{0} + k \delta t), \omega_{0} (t_{0} + k \delta t)), (\bm{m}^{(k)}(t_{0} + k \delta t), \omega_{0} (t_{0} + k \delta t + \Delta t_{k}))]}{\varepsilon}
\\
  &=
  \frac{ \sqrt{ (\ell^{(k)\prime}[\bm{m}(t_{0} + k \delta t),\bm{m}^{(1)}(t_{0}+ k \delta t)])^{2} + (\omega_{0}\Delta t_{k})^{2} }}{\varepsilon}, 
\end{split}
\end{equation}
\end{widetext}
where $\ell^{(k)\prime}[\bm{m}(t_{0}+k \delta t),\bm{m}^{(1)}(t_{0}+k \delta t)]=\cos^{-1}[\bm{m}(t_{0} + k \delta t)\cdot\bm{m}^{(k)}(t_{0} + k \delta t)]$. 
The perturbation to the time variable in the oscillating signal obeys the recurrence formula, 
\begin{equation}
  \omega_{0}
  \Delta t_{k+1}
  =
  \frac{\Delta t_{k}}{p_{k}},
\end{equation}
whereas $\bm{m}^{(k+1)}(t_{0}+k \delta t)$ is defined by rotating $\bm{m}(t_{0}+k \delta t)$ to the direction of $\bm{m}^{(k)}(t_{0}+k \delta t)$ 
with an angle $\ell^{(k+1)}$ given by 
\begin{equation}
  \ell^{(k+1)}
  =
  \frac{\ell^{(k) \prime}}{p_{k}}. 
\end{equation}
Repeating the procedure, 
the Lyapunov exponent is obtained from Eq. (\ref{eq:def_lambda}). 



\section{Magnitudes of alternative current and microwave field}
\label{sec:AppendixD}

In the main text, we compared, for example, the locking ranges of the forced synchronization caused by the alternating current and microwave field. 
The comparison should be discussed on an equal basis. 
In this Appendix, we discuss the correspondence between the magnitudes of the alternating current and the microwave field used in Secs. \ref{sec:Response to alternating current} and \ref{sec:Response to microwave field}. 
The magnitude of the alternating current used in Sec. \ref{sec:Response to alternating current} is on the order of $1$ mA at maximum 
for the investigation of the forced synchronization. 
The same amount of current generates a microwave field with the amplitude of $j_{\rm ac}/(2\pi R)$, 
where $R$ is the distance from the current source and the STO. 
Assuming that the microwave source is placed on the STO with the distance of $10$ nm, 
the amplitude of the microwave field is estimated to be $200$ Oe. 
Therefore, it is reasonable to assume that $H_{\rm ac}$ in Sec. \ref{sec:Response to microwave field} is on the order of $100$ Oe. 


The microwave field is experimentally generated by applying the alternating current to a coplanar wave guide. 
We understand that it is experimentally difficult to place a coplanar wave guide within a such short distance from the STO 
because there are many layers on the free layer, such as upper lead and MgO \cite{kubota13}. 
One might consider, for example, that the upper lead is substituted as a microwave source, 
which enables us to generate the microwave close to the free layer. 
Considering these points, we consider that the above estimation of the microwave amplitude is an upper limit of the experimentally available value in a conventional technology. 
Simultaneously, we emphasize that, for example, the $H_{\rm ac}=1.0$ Oe used in Fig. \ref{fig:fig9} is,a reasonable value in experiments \cite{tsunegi19}. 


We also note that the fieldlike torque due to the alternating current plays the same role with the torque due to the linearly polarized microwave field. 
The effective field of the fieldlike torque is estimated to be 
$\beta \hbar \eta j_{\rm ac}/(2eMV)$, where $\beta$ is the dimensionless parameter 
corresponding to the ratio of the fieldlike torque to the Slonczewski torque. 
In a metallic system, $\beta$ is roughly $g_{\rm i}^{\uparrow\downarrow}/g_{\rm r}^{\uparrow\downarrow}$, where 
$g_{\rm r}^{\uparrow\downarrow}$ and $g_{\rm i}^{\uparrow\downarrow}$ are the real and imaginary parts of mixing conductance \cite{kovalev02}. 
In a magnetic tunnel junction, $\beta$ is related to the tunneling probability at the interface \cite{theodonis06}. 
It has been found that $|\beta|<1$ in both cases. 
Then, assuming that $j_{\rm ac}=1$ mA and $|\beta|=0.1$, the effective field of the fieldlike torque is estimated to be $|\beta \hbar \eta j_{\rm ac}/(2eMV)|\simeq 5.3$ Oe. 
The value is on the same order with that studied in the main text; see, for example, Fig. \ref{fig:fig8}. 
Therefore, the role of the fieldlike torque on the synchronization can be understood from the results shown in the main text. 



%



\end{document}